\documentclass[a4paper,11pt]{article}
\usepackage{jheppub} 
\usepackage[T1]{fontenc} 
\usepackage{epsfig}
\usepackage{cancel,slashed}
\usepackage{multirow}
\preprint{\\CERN-PH-TH/2012-353, IPHC-PHENO-12-08, PCCF-R1-12-08}

\title{Searching for sgluons in multitop events at a center-of-mass energy of 
8 TeV}
\author[a]{Samuel Calvet,}
\author[b,c]{Benjamin Fuks,}
\author[a]{Philippe Gris,}
\author[a]{Lo\"ic Val\'ery}

\affiliation[a]{Laboratoire de Physique Corpusculaire, Clermont Universit\'e / Universit\'e Blaise Pascal / CNRS-IN2P3, 24 avenue des Landais, F-63177 Aubi\`ere Cedex, France}
\affiliation[b]{Theory Division, Physics Department, CERN, CH-1211 Geneva 23,
  Switzerland}
\affiliation[c]{Institut Pluridisciplinaire Hubert Curien/D\'epartement
  Recherches Subatomiques, Universit\'e de Strasbourg/CNRS-IN2P3,
  23 rue du Loess, F-67037 Strasbourg, France}
\emailAdd{scalvet@in2p3.fr}
\emailAdd{fuks@cern.ch}
\emailAdd{gris@clermont.in2p3.fr}
\emailAdd{valery@clermont.in2p3.fr}

\abstract{
Large classes of new physics theories predict the existence of new scalar
states, commonly dubbed sgluons, lying in the adjoint representation of the QCD
gauge group. Since these
new fields are expected to decay into colored Standard Model particles, and in
particular into one or two top quarks, these theories predict a possible
enhancement of the hadroproduction rate associated with multitop
final states. We therefore investigate 
multitop events produced at the Large Hadron Collider, running at a 
center-of-mass energy of 8 TeV, and employ those events to probe the
possible existence of color adjoint scalar particles. We first
construct a simplified effective field theory motivated by $R$-symmetric
supersymmetric models where sgluon fields decay dominantly into top quarks.
We then use this model to analyze  
the sensitivity of the Large Hadron Collider in both a multilepton plus jets
and a single lepton plus jets channel. After having based our event
selection strategy on the possible presence of two, three and four top quarks in
the final
state, we find that sgluon-induced new physics contributions to multitop cross
sections as low as 10-100 fb can be excluded at the 95\% confidence level,
assuming an integrated luminosity of 20 fb$^{-1}$. Equivalently, sgluon masses
of about 500-700 GeV can be reached for several classes of benchmark scenarios.
}


\begin{document}

\def\be{\begin{equation}}
\def\ee{\end{equation}}
\def\bea{\begin{eqnarray}}
\def\eea{\end{eqnarray}}
\def\bsp#1\esp{\begin{split}#1\end{split}} 
\def\lag{{\cal L}}


\maketitle
\flushbottom


\section{Introduction}
The Standard Model of particle physics has passed for many years all
experimental tests. Only the mechanism of electroweak symmetry breaking, which
is currently being addressed by the general-purpose experiments ATLAS and CMS of
the Large Hadron Collider (LHC), at CERN, 
has remained for a long time an unsolved question.  The recent observation of a
neutral bosonic particle compatible with a Standard-Model-like Higgs boson  
\cite{Atlas:2012gk, Chatrchyan:2012gu}, together with the imminent
measurement of its properties, would then represent an impressive success of
this theoretical framework. However, the mass of a
fundamental scalar field is drastically affected by quantum corrections,
which leads to the conceptual question of its stabilization with respect to the
Planck scale lying orders of magnitude away from the weak scale. Over the last
decades, large classes of alternative theories have been
proposed in order to extend the Standard Model and cure this issue. 
Among these, weak-scale
supersymmetry, and in particular its minimal version, the Minimal Supersymmetric
Standard Model (MSSM) \cite{Nilles:1983ge, Haber:1984rc}, is one of the most
theoretically and experimentally studied option. It is 
known to solve this so-called ``hierarchy problem'' by introducing partners to
the Standard Model degrees of freedom with opposite statistics. In addition,
several other conceptual problems of the Standard Model are addressed, such as
the unification of the gauge coupling strengths at high energies or the
question of a viable dark matter candidate. 

Experimental searches for the supersymmetric partners of the Standard Model
particles are therefore among the main topics investigated at the LHC.
Up to now, both the ATLAS and CMS collaborations have mainly focused on the
strong production channels, since they yield larger cross sections. As a
result, no traces of squarks and gluinos have been detected so far
and the limits on the masses of the
first-generation and second-generation squarks as well as those on the mass of
the gluino are consequently pushed to a higher and higher scale
\cite{Lowette:2012uh, atlassusy, cmssusy}. Therefore, the
experimental attention starts to shift towards third-generation squarks and
electroweak production
channels. However, all the current results may not be valid for more general
supersymmetric scenarios. They are indeed either derived in
the framework of the constrained version of the MSSM, where the hundreds of
free parameters of the general model are reduced to a set of four parameters and
one sign, or in the context of simplified models inspired by the same
constrained MSSM. In contrast,
there are a vast variety of non-minimal supersymmetric models which are
valuable to be investigated at the LHC. In particular, some of the final state
signatures predicted by these non-minimal models require dedicated
phenomenological studies in order to be ready for the interpretation of the data
in these theories. 
This is one of the scopes of the work presented in this paper.

Motivated by these arguments, we address the production and decay
of color-octet massive scalar particles, also dubbed sgluons, that are predicted
by several non-minimal supersymmetric models. The most
considered examples are $N\!=\!1\!/\!N\!=\!2$ hybrid
supersymmetric theories \cite{Fayet:1975yi, AlvarezGaume:1996mv, Plehn:2008ae,
Choi:2008pi, Choi:2008ub, Choi:2009jc, choi:2010gc, Choi:2010an,
Schumann:2011ji, Kotlarski:2011zz, Kotlarski:2011zza}
and $R$-symmetric supersymmetric theories \cite{Salam:1974xa,
Fayet:1974pd, Kribs:2007ac}. In these models, the vector supermultiplet of the
MSSM associated to the QCD gauge group is supplemented by an additional chiral
supermultiplet lying in the adjoint representation of $SU(3)$. 
This new supermultiplet contains, on the one hand, a two-component fermionic
component which
mixes with the usual gluino to form a four-component Dirac fermion. On the
other hand, it also includes a color-octet complex scalar particle,
\textit{i.e.}, a sgluon field. Let us note that 
color octet scalar particles also appear in vector-like confining theories
\cite{Kilic:2008pm, Kilic:2008ub, Kilic:2009mi, Kilic:2010et,
Dicus:2010bm, Sayre:2011ed} or in extra-dimensional models \cite{Burdman:2006gy}. 

As all of these models are similar in the sense of the LHC signatures, 
we adopt the approach introduced by the LHC New Physics Working Group
\cite{Alves:2011wf} and employ a simplified model describing a scalar octet
field and its interactions with the Standard Model sector
\cite{Brooijmans:2012yi}. This has the major advantages to leave open the
possibility of reinterpreting the results in the context of 
any of the original models (or even in the framework of another theory including
a sgluon field) and also to avoid handling
a complete model which demands a careful design of a theoretically motivated
but not experimentally excluded benchmark scenario. Equivalently, rather than
fixing hundreds of free parameters related to one of the new physics
theories above-mentioned, we focus
on a specific sector of the model, relevant for our study, that is
described by a small number of couplings and masses.

We therefore start by constructing a simplified model describing the dynamics of
a scalar field lying in the
adjoint representation of the QCD gauge group. Motivated by complete models
where loop-induced operators imply sgluon fields singly coupled to
Standard Model quarks and gluons, we also include the corresponding interactions
in our effective theory.
Subsequently, once produced through standard strong interactions, a sgluon
can then decay either to a quark
pair or to a gluon pair, the latter being in general dominant for a sgluon field
with a rather low mass \cite{Plehn:2008ae,Choi:2008ub}. If kinematically
allowed, the same interactions ensure a possible decay to a pair of top
quarks or to an associated pair comprised of one single top quark and a light jet. 

In this scenario, sgluon pair-production and decay at the
LHC could lead to signatures possibly containing two, three or even four top
quarks, the last two final states being expected to be largely suppressed in the
context of the Standard Model. In this paper, we perform a phenomenological
analysis of the sensitivity
of the LHC collider, running at a center-of-mass energy $\sqrt{s}$ of 8 TeV, to
the possible observation of sgluon fields through multitop signatures.
This work extends our previous contribution to the 2011 Les Houches workshop 
\cite{Brooijmans:2012yi} by including a more accurate description of the
Standard Model backgrounds and a more efficient search strategy. It
is also
complementary to recent ATLAS and CMS investigations where sgluon-induced 
four-jet signatures have been considered \cite{Aad:2011yh, Aad:2011fq, ATLAS:2012bi,
CMS:2011bi}. In these analyses, limits on the sgluon mass up to 2 TeV have been
extracted. In our theoretical setup, we evade these bounds by imposing the
sgluon field to dominantly decay to at least one top quark, as motivated by
realistic supersymmetric models including a color-adjoint scalar particle.
Consequently, lower sgluon masses can be possibly expected.

This paper is organized as follows. In Section \ref{sec:theo}, we describe our
proposal for a simplified theory modeling sgluon pair production and decay at
the LHC and define
benchmark scenarios for our phenomenological studies. For each benchmark point,
we present the sgluon decay table and the leading-order and next-to-leading
total production cross sections. Section
\ref{sec:technical} is dedicated to our technical setup for
the Monte Carlo simulation of both signal and background events at $\sqrt{s} =
8$ TeV. A
particular emphasis is put on the fast detector simulation package that we have 
used and in the way objects are reconstructed. In Section
\ref{sec:analysis}, we present the details of our phenomenological analyses and
the associated results. Our conclusions are given in Section
\ref{sec:conclusions}.

\section{A simplified model for sgluon production and decays at the
LHC}\label{sec:theo}
In order to investigate sgluon production and decay at the LHC, we
construct a simplified model describing sgluon interactions
with the Standard Model fields. To this aim, we follow the approach of Ref.\
\cite{Alves:2011wf} and extend the Standard Model in a minimal way by supplementing to
its particle content one massive real scalar field $\sigma$ lying in the adjoint
representation of the QCD gauge group. Its kinetic and mass terms are standard
and can be expressed in terms of the QCD covariant derivative taken in the
adjoint representation,
\be\label{eq:lkin}
  \lag_{\rm kin} = \frac12 D_\mu \sigma^a D^\mu \sigma_a -
     \frac12 m_\sigma^2  \sigma^a \sigma_a
  \qquad\text{with}\qquad  
  D_\mu \sigma^a = \partial_\mu \sigma^a + g_s\ f_{bc}{}^a\ G_\mu^b\ \sigma^c \
  ,
\ee
where adjoint gauge indices are explicitly indicated. 
In the expressions above, we have introduced the strong coupling constant $g_s$,
the antisymmetric structure constants of $SU(3)$ $f_{bc}{}^a$, the
gluon field $G_\mu^b$ and the sgluon mass $m_\sigma$.

These interactions do not include single sgluon couplings to Standard Model
fields. However, the presence of additional
particles in general induces, in complete theories, loop diagrams leading to
effective operators describing 
the coupling of a single sgluon field to up-type ($u$) and down-type ($d$)
quark pairs, as well as to gluon pairs. In our theoretical framework, those
interactions are described by the effective Lagrangian
\be
 \lag_{\rm eff} = 
   \sigma^a \bar d T_a \Big[ a^L_d P_L + a^R_d P_R \Big] d + 
   \sigma^a \bar u T_a \Big[ a^L_u P_L + a^R_u P_R \Big] u + 
   a_g d_a{}^{bc} \sigma^a G_{\mu\nu b} G^{\mu\nu}{}_c + {\rm h.c.}  \ .
\label{eq:leff}\ee
The matrices $T_a$ and the tensor $d_a{}^{bc}$ are respectively the fundamental 
representation matrices and the symmetric structure constants of $SU(3)$, the
operators $P_L$ and $P_R$ are the left-handed and right-handed chirality
projectors acting on spin space and $G_{\mu\nu}{}^a$ is the gluon field strength
tensor. We have also introduced the parameters $a_q^L$ and $a_q^R$, with
$q=u,d$, to model the strengths of the interactions among left-handed and
right-handed quarks $q$ and a single sgluon, respectively, as well as one 
single parameter $a_g$ for the modeling of the
interactions among two gluons and one sgluon.
It is important to note that the interactions included in the Lagrangian of Eq.\
\eqref{eq:leff} open all the
possible sgluon decays to Standard Model colored particles.

Inspecting the two Lagrangians of Eq.\ \eqref{eq:lkin} and Eq.\ \eqref{eq:leff},
one observes that our simplified theory is described by one mass parameter, the
sgluon mass $m_\sigma$, and the effective couplings of sgluons to colored
partons described by a set of four complex $3\times 3$ matrices in flavor space
$a^L_d$, $a^R_d$, $a^L_u$ and $a^R_u$, together with one real (dimensionful) 
number $a_g$.
Assuming ${\cal O}(1)$ effective interactions, sgluon masses up to about 2
TeV are already excluded by dijet resonance searches \cite{Aad:2011fq}.
To evade this constraint, we choose
scenarios where the sgluon field dominantly decays into final states containing 
at least one top quark or where its couplings to a pair of light quarks or to a
pair of  gluons are reduced. This choice is motivated by $R$-symmetric
supersymmetric theories and by $N\!=\!1\!/\!N\!=\!2$ hybrid supersymmetric models.
In these models, interactions among a single sgluon and a
pair of quarks are driven by loops of squarks and gluinos (and thus suppressed by 
their heavy propagators). The computation of
those loops lead to non-vanishing effective couplings only if at 
least one of the external quarks is a top quark. 
Furthermore, the coupling strength $a_g$ is related to a dimension-five operator
and is thus expected to be suppressed too. 

We therefore consider two series of benchmark
scenarios, refered to as scenarios of class I and class II in the following. 
For the first set of scenarios, sgluon particles are allowed to
decay, in a universal way, to any associated pair of up-type quarks containing
at least one top quark. We subsequently fix
\be\label{bchIa}
  (a^L_u)^3{}_i = (a^R_u)^3{}_i =  (a^L_u)^i{}_3 = (a^R_u)^i{}_3 = 3 \cdot 10^{-3} \ ,
\ee
for $i=1,2,3$ and impose that all other interactions among quarks and sgluons
are vanishing. 
From the explicit calculations of the relevant loop diagrams in Ref.~\cite{Plehn:2008ae},
the choice of Eq.~\eqref{bchIa} corresponds to a 
scenario where squarks and gluinos have typical masses of about 1-2~TeV and that allows for
non-minimal flavor violation in the squark sector induced by supersymmetry breaking.
Concerning the parameter $a_g$ related to the strength of the coupling among
sgluons and gluons, we choose the value 
\be\label{bchIb}
  a_g = 1.5\times 10^{-6}  \text{ GeV}^{-1}\ .
\ee
which arises in realistic scenarios with supersymmetric masses of the same order 
as above and when left-handed and right-handed squarks are almost, but not, 
mass-degenerate \cite{Plehn:2008ae}.

\begin{table}[t]
\begin{center}
\begin{tabular}{| c || c  c |}
\hline
Parameters &  Scenarios of type I & Scenarios of type II\\
\hline
$a_g$ & $1.5 \times 10^{-6}$ GeV$^{-1}$ & $1.5 \times 10^{-6}$ GeV$^{-1}$ \\
$(a_u)^3{}_3$ & $3\cdot 10^{-3}$ & $3\cdot 10^{-3}$\\
$(a_u)^3{}_1 = (a_u)^1{}_3$ & $3\cdot 10^{-3}$ & 0 \\
$(a_u)^3{}_2 = (a_u)^2{}_3$ & $3\cdot 10^{-3}$ & 0 \\
$m_\sigma$ & [200-1000] GeV & [400-1000] GeV \\
$m_t$ & 172 GeV & 172 GeV\\
 \hline
\end{tabular}
\caption{\label{tab:params} Non-zero input parameters for benchmark scenarios of
class I (second column) and II (third column). All the Standard Model parameters
but the mass of the top quark (indicated in the Table) follow the conventions of
Ref.\ \cite{Christensen:2009jx}.}
\end{center}
\end{table}

In scenarios of class II, we focus exclusively on sgluon-induced LHC signatures
with four top quarks and therefore forbid flavor violation in the sgluon decays.
The only non-vanishing effective interactions are thus driven by the parameters
\be
  (a^L_u)^3{}_3 =  (a^R_u)^3{}_3 = 3\cdot 10^{-3}
  \qquad\text{and}\qquad
  a_g = 1.5\times 10^{-6} \text{ GeV}^{-1} \ .
\ee

The values of the non-zero parameters of the Lagrangians included in Eq.\
\eqref{eq:lkin} and Eq.\ \eqref{eq:leff} are summarized, for the two series of
benchmark points, in Table \ref{tab:params}.
For all the Standard Model parameters, but the top quark mass $m_t$ which is
chosen equal to 172 GeV, we follow the conventions of Ref.\
\cite{Christensen:2009jx}.
Moreover, the sgluon mass is kept free and allowed to vary in the range
$m_\sigma\in [200 - 1000]$ GeV and $m_\sigma\in [400 - 1000]$ GeV for 
class I and class II scenarios, respectively.

A key element in the multitop analysis of sgluon production and decay at the LHC
lies in the sgluon branching fraction to final states containing one or 
two top quarks. To study the evolution of these branching ratios with the
sgluon mass, the Lagrangians of Eq.\ \eqref{eq:lkin} and Eq.\ \eqref{eq:leff}
have been implemented in the {\sc FeynRules} package
\cite{Christensen:2008py, Christensen:2009jx, Christensen:2010wz, Duhr:2011se,
Fuks:2012im} and
the model has been exported to the UFO format \cite{Degrande:2011ua}.  We have
subsequently used the matrix-element generator 
{\sc MadGraph} 5 \cite{Alwall:2011uj} to compute all
sgluon partial decay widths. We then estimate the total width and the
different branching ratios into two gluons,
into an associated pair of a top quark and a light quark and into two top quarks.
The results are shown in Table \ref{tab:sigma} for the two classes of considered
scenarios. 

The branching of a light sgluon of few hundreds of GeV 
into a top-antitop pair is, as expected, kinematically suppressed compared to
the other decay channels. This also holds for
scenarios of class II where the sgluon field most of the time decays into a pair
of gluons when it is light. For both classes of scenarios, the branching ratio
of the $t\bar t$ decay increases with the sgluon mass (if kinematically allowed). 
However, the
contributions of the dijet channel to the total width also become more
important so that the branching into a top-antitop
pair peaks for $m_\sigma\sim 800$ GeV and $m_\sigma\sim 600$ GeV for
scenarios of class I and II, respectively, and then decreases for heavier
sgluons.

\begin{table}[t]
\begin{center}
\begin{tabular}{|c|| c c || c c c || c c|}
\hline
  Scenario & $m_\sigma$ [GeV] & $\Gamma_\sigma$ [MeV] & $BR(t\bar t)$ &
  $BR(tj / \bar t j)$ & $BR(gg)$ & $\sigma_{\rm tot}$ [fb] & $K_{\rm
  NLO}$\\
\hline
  I & 200 & 0.012& - & 80\% & 20\% & 98600 & 1.6\\
\hline
  I & 300 & 0.105 & - & 92.3\% & 7.7\% & 9802  & 1.6  \\
\hline
  I & \multirow{2}{*}{400} & 0.219 & 4.4 \% & 86.9\% & 8.7\%  &
    \multirow{2}{*}{1625} & \multirow{2}{*}{1.7}\\
  II&                      & 0.029& 33.3\% & -      & 66.7\% & & \\
\hline
  I & \multirow{2}{*}{500} & 0.350 & 9.8 \% & 79.5\% & 10.1\%  &
    \multirow{2}{*}{358.1} & \multirow{2}{*}{1.8}\\
  II&                      & 0.072& 47.8\% & -      & 52.2\% & & \\
\hline
  I & \multirow{2}{*}{600} & 0.485 & 12 \% & 75\% & 13\%  &
    \multirow{2}{*}{94.9} & \multirow{2}{*}{1.8}\\
  II&                      & 0.124& 48\% & -      & 52\% & & \\
\hline
  I & \multirow{2}{*}{700} & 0.628 & 13.2 \% & 70.5\% & 16.3\%  &
    \multirow{2}{*}{28.4} & \multirow{2}{*}{1.9}\\
  II&                      & 0.185& 44.7\% & -      & 55.3\% & & \\
\hline
  I & \multirow{2}{*}{800} & 0.779 & 13.5 \% & 66.9\% & 19.6\%  &
    \multirow{2}{*}{9.26} & \multirow{2}{*}{2.0}\\
  II&                      & 0.252& 41\% & -      & 59\% & & \\
\hline
  I & \multirow{2}{*}{900} & 0.943 & 13.5 \% & 63.4\% & 23.1\%  &
    \multirow{2}{*}{3.22} & \multirow{2}{*}{2.1}\\
  II&                      & 0.345 & 36.9\% & -      & 63.1\% & & \\
\hline
  I & \multirow{2}{*}{1000} & 1.120 & 13.2 \% & 60.2\% & 26.6\%  &
    \multirow{2}{*}{1.17} & \multirow{2}{*}{2.2}\\
  II&                      & 0.447& 33.2\% & -      & 66.8\% & & \\
\hline
\end{tabular}
\caption{\label{tab:sigma} Dependence on the sgluon mass $m_\sigma$ of the
  sgluon total width ($\Gamma_\sigma$), of the different branching ratios ($BR$)
  to Standard Model colored particles and of the total cross section at leading
  order ($\sigma_{\rm tot}$). The next-to-leading order $K$-factors $K_{\rm NLO
  }$ are also indicated, extracted from Ref.\ \cite{GoncalvesNetto:2012nt}. In
  the notations of the table, the symbol $j$ stands for an up or charm quark
  while $g$ refers to a gluon.}
\end{center}
\end{table}

\begin{figure}
  \begin{center}
    \includegraphics[width=\columnwidth]{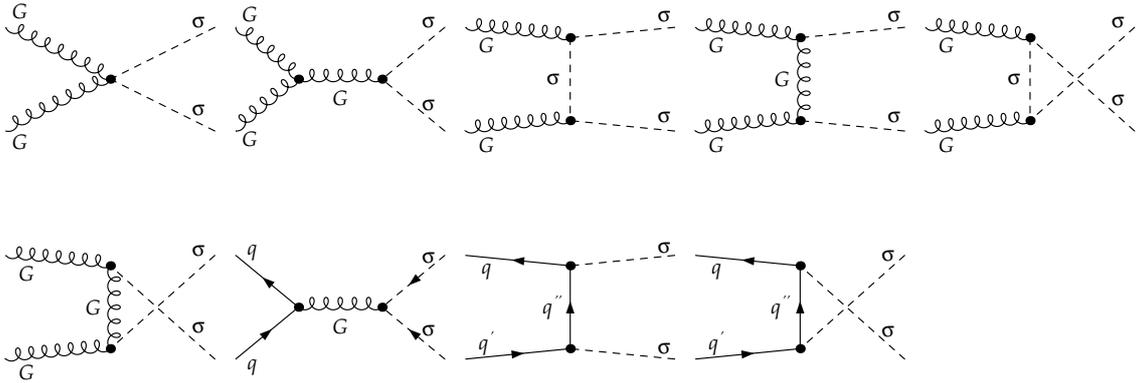}
    \caption{Tree-level Feynman diagrams associated with sgluon pair production
    at hadron colliders. These diagrams correspond to the interactions included
    in the Lagrangians of Eq.\ \eqref{eq:lkin} and Eq. \eqref{eq:leff} and have
    been created by means of the program {\sc FeynArts} \cite{Hahn:2000kx}.}
    \label{fig:diag}
  \end{center}
\end{figure}

We also show in Table \ref{tab:sigma} the leading-order (LO) sgluon production
cross sections as computed by the {\sc MadGraph} 5 program for the LHC collider
running at a center-of-mass energy of 8 TeV. The presented results correspond to
the convolution of the tree-level matrix elements related to the Feynman
diagrams of 
Figure \ref{fig:diag} with the LO set of the CTEQ6 parton density fit
\cite{Pumplin:2002vw} and renormalization and factorization scales fixed
to the transverse mass of the produced heavy particles. As sgluon-pair
production cross sections are known at the next-to-leading order (NLO) accuracy
within the {\sc MadGolem} setup \cite{GoncalvesNetto:2012nt}, we also indicate
in Table \ref{tab:sigma} the corresponding NLO $K$-factors. In our
phenomenological analyses, signal event samples are then normalized
according to the NLO results.
 
\section{Technical setup for the Monte Carlo simulations}
\label{sec:technical}

\subsection{Objects definitions and detector simulation}
\label{sec:object}

\begin{figure}
  \begin{center}
    \includegraphics[width=0.65\columnwidth]{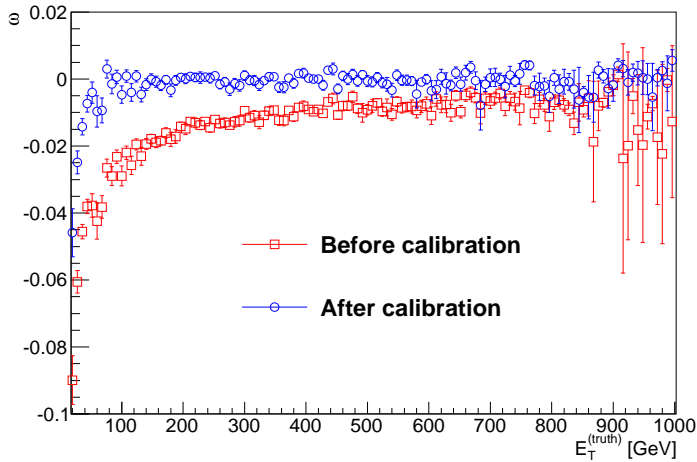}
    \caption{Evolution of the $\omega$-variable defined in Eq.\ \eqref{eq:omega} 
       with respect to the true transverse energy of the reconstructed
       jet before (red squares) and after (blue circles) calibration.}
    \label{fig:JES_omega}
  \end{center}
\end{figure}

Detector simulation is performed with the {\sc Delphes}
program, using the public ATLAS card \cite{Ovyn:2009tx}. Jets are hence
reconstructed by means of an anti-$k_{t}$ algorithm with a radius parameter set to 
$R=0.4$, as provided by the {\sc FastJet} package \cite{Cacciari:2005hq}. 
However, jets defined with such a low radius, and in particular those with a
small transverse momentum, may suffer from large bias in energy reconstruction 
due to the magnetic field as simulated by {\sc Delphes},
which spreads out the energy within the detector. Such an effect may be
described by the variable
\be
  \omega = \frac{E_T^{\text{(reco)}} - E_T^\text{(truth)}}{E_T^\text{(truth)}} \
  ,
\label{eq:omega}\ee
where $E_T^{\text{(reco)}}$ is the reconstructed jet transverse energy and
$E_T^{\text{(truth)}}$ is the true transverse jet energy defined as the jet
transverse energy before detector simulation. The evolution of the
$\omega$-variable with the (true) jet energy is presented on Figure
\ref{fig:JES_omega} (red squares) in the context of dijet events issued from the
decay of a sequential $Z'$-boson whose mass varies in the range [200, 1000] GeV.
The energy loss reaches
about $5 \%$ for jets with $E_T^{\text{(truth)}} = 20$~GeV while it stabilizes
at about $1 \%$ for jets with a transverse energy $E_T^{\text{(truth)}} > 500$ GeV.

To account for this effect, an \textit{ad-hoc} calibration is estimated from the
above-mentioned dijet events. We apply on the reconstructed jet energy 
the correction function 
\be
  E_T^{\rm (cal)} = E_T^{\rm (reco)} \times \left[ 2.62 \cdot 10^{-3} - 
    \frac{ 0.451{\rm GeV}}{E_T^{\rm (reco)}}  \ln \frac{E_T^{\rm (reco)}}{{\rm
    GeV}} \right] ,  
\ee 
where $E_T^{\rm (cal)}$ is the jet transverse energy after calibration and we
show the associated effects on Figure \ref{fig:JES_omega} (blue circles). This
calibration procedure allows us to recover the correct jet energy for transverse
energy as low as $E_T^{\text{(truth)}}$ of about 40 GeV.

In our analysis, only jets with a transverse energy $E_T > 20$ GeV (after
calibration) and a pseudorapidity $|\eta|<2.5$ are retained.
In addition, we estimate a $b$-tagging efficiency of about 60\%, together with a
charm and light flavor mistagging rate of 10\% and 1\%, respectively.

Charged leptons candidates\footnote{By the generic terminology 
\textit{charged leptons}, we only consider electrons and muons.} 
are requested to have a transverse momentum $p_T$ larger than 20 GeV and  
a pseudorapidity $|\eta| < 2.47$ and $|\eta| < 2.5$ for electrons and muons,
respectively. We also impose
two isolation criteria. First, the closest jet to an electron is
removed if the angular distance $\Delta R  = \sqrt{\Delta\phi^2 + \Delta\eta^2}
\leq 0.1$, where $\phi$ stands for the azimutal angle with respect to the beam
direction. Secondly, in
the case at least one jet is present within a cone of radius $R=0.4$
centered on the lepton, the lepton is removed from the event.

\subsection{Monte Carlo simulations of the signal and background processes}
\label{sec:mc}
We perform in this work a prospective phenomenological analysis of
multitop events induced by the production of a pair of sgluon fields. We aim to
estimate the LHC sensitivity to the presence of sgluons in such events
by means of a phenomenological analysis employing
Monte Carlo simulations. Our analysis context consists of the
LHC collider running at
a center-of-mass energy of $\sqrt{s}=8$ TeV and we normalize our event samples
to an integrated luminosity of 20 fb$^{-1}$, which corresponds
to the expectations for the end of the 2012 LHC run \cite{chamonix}. 

The hard scattering processes related to both the
signal and the different sources of background have been described with the
matrix-element generator
{\sc MadGraph} 5 \cite{Alwall:2011uj}. Using the QCD factorization theorem, the matrix elements
are convolved with the leading order set of the CTEQ6 parton
density fit \cite{Pumplin:2002vw} and the renormalization and
factorization scales are fixed to the transverse mass of the produced heavy
particles. Parton-level events are then integrated into a full hadronic
environment by matching the hard scattering matrix elements to the full parton 
showering and hadronization infrastructure as provided by the {\sc Pythia} 6
package \cite{Sjostrand:2006za}. We have then included a fast detector
simulation as performed by the program {\sc Delphes} 1.9 \cite{Ovyn:2009tx},
using, as stated above, the publicly available ATLAS detector card. 

The procedure above is however known not to provide
an accurate description of the kinematical properties of jets due to an
underestimation of the hard emission effects by parton showering
algorithms which only correctly model
jets in the soft and collinear limit. In contrast, matrix elements properly 
describe additional hard and widely separated radiation, but are known to break
down in the soft and collinear kinematical regions. 
Therefore, we allow for the matrix elements related to our 
background samples to contain zero, one, two, \textit{etc}, additional
hard jets. The resulting events are then merged following the Mangano (MLM)
procedure  \cite{Mangano:2006rw} as implemented in the {\sc MadEvent}
program \cite{Alwall:2008qv}. 

In this setup, two parton-level selection criteria are imposed. Firstly, the squared 
jet measure $k_T^2= 
\min(p_{Ti}^2, p_{Tj}^2) R_{ij}$ related to two final state partons is asked to
be larger than a process-dependent value $(k_T^{\rm min})^2$, where  $p_{Ti}$
and $p_{Tj}$ are the transverse momenta of the two partons and $R_{ij}$ their
angular distance in the $(\eta,\phi)$ plane. Secondly, the jet measure related
to a final-state and an initial-state parton is defined as the transverse
momentum of the final-state parton and is asked to be larger than a
process-dependent value $p_T^{\rm min}$.
The events are then passed to {\sc Pythia} and jets are reconstructed 
using the {\sc FastJet} program \cite{Cacciari:2005hq}, making use of a 
$k_T$-jet algorithm with a cut-off scale $Q^{\rm m}$. 
An event is kept only if, for each of the reconstructed jets in the event, the
jet measure between the jet and the parton which it
is originating from is smaller than $Q^{\rm m}$. 
In order to maintain the full inclusiveness of the merged sample, extra jets
with respect to the original number of partons are allowed in the sample with
the highest
multiplicity. The merging scale $Q^{\rm m}$ is process-dependent and
chosen in such a way that the differential jet rate distributions of the merged
samples are smooth. The values chosen for the parton-level selections, the
maximum number of included hard emissions and the merging scale are given in
Table \ref{tab:xsec} for the various background processes. Concerning the
signal, no merging with matrix elements of a higher multiplicity has been
performed as the lowest order subprocess already contains many hard jets.

\begin{table}[t]
\begin{center}
\begin{tabular}{| l || c c || c c || c c |}
\hline
Process & $k_T^{\rm min}$ [GeV] & $p_T^{\rm min}$ [GeV] & $n$ &
$Q^{\rm m}$ [GeV]& $\sigma$ [pb]& $N$\\
\hline
  $W(\to 1\ell) + \text{jets}$ & 10 & 10  & 4 & 20 & 35678    & $2\cdot
    10^{10}$\\
  $\gamma^{*}/Z(\to 2 \ell) + \text{jets}$ & 10 & 10 & 4 & 20 & 3460  & $4
\cdot 10^6$\\
\hline 
  $t\bar{t} (\to 1 \ell) + \text{jets}$& 20 & 20 & 2 & 30 & 112.& $9\cdot 10^6$
   \\
  $t\bar{t} (\to 2 \ell) + \text{jets}$& 20 & 20 & 2 & 30 & 27.2& $3\cdot 10^6$
    \\
\hline 
  $t/\bar t+ \text{jets}$ [$t$, incl.] & - & - & 0 & - & 28.4 & $4\cdot10^6$ \\
  $t/\bar t+ \text{jets}$ [$tW$, incl.]& - & - & 0 & - & 12.1 & $2\cdot10^6$ \\
  $t/\bar t+ \text{jets}$ [$s$, incl.] & - & - & 0 & - & 1.81 & $8\cdot10^5$ \\
\hline
  $WW(\to 1\ell)+ \text{jets}$ &10 & 10 & 2& 20 & 24.3 & $3\cdot10^6$\\
  $WW(\to 2\ell)+ \text{jets}$ &10 & 10 & 2& 20 & 5.87 & $8\cdot10^5$ \\
  $WZ(\to 1\ell)+ \text{jets}$ &10 & 10 & 2& 20 & 6.47 & $2\cdot10^5$ \\
  $WZ(\to 2\ell)+ \text{jets}$ &10 & 10 & 2& 20 & 1.58 & $2\cdot10^5$ \\
  $WZ(\to 3\ell)+ \text{jets}$ &10 & 10 & 2& 20 & 0.76 & $7\cdot10^4$ \\
  $ZZ(\to 4\ell)+ \text{jets}$ &10 & 10 & 2& 20 & 0.17 & $4\cdot10^4$ \\
  $ZZ(\to 2\ell)+ \text{jets}$ &10 & 10 & 2& 20 & 1.50 & $4\cdot10^4$ \\
\hline 
  $t\bar{t}W  + \text{jets}$ [incl.] & 10& 10 & 2 & 20  & 0.25   &
    $3\cdot10^4$\\
  $t\bar{t}Z  + \text{jets}$ [incl.] & 10& 10 & 2 & 20  & 0.21   & 
    $5\cdot10^4$\\
  $t\bar{t}WW + \text{jets}$ [incl.] & 10& 10 & 2 & 20  & 0.013  & 
    $2\cdot10^3$\\
  $t\bar{t}t\bar{t} + \text{jets}$ [incl.]& - & - & 0 & - & $7 \cdot 10^{-4}$& $10^{3}$ \\
\hline
\end{tabular}
\caption{We present the simulated background processes, together with the set of
applied parton-level selection criteria ($k_T^{\rm min}$ and $p_T^{\rm min}$),
the number of 
allowed extra hard emissions at the matrix-element level ($n$) and the matching
scale ($Q^{\rm m}$). The numerical values employed for the cross sections
($\sigma$) are
also given, together with the number of generated events ($N$). For each of the
background process, the final state contains at least one lepton $\ell$, where
$\ell$ stands equivalently for electrons, muons, leptonic
and hadronic taus. The notations \textit{incl.} indicates that the associated
samples are inclusive in the decays. We refer to the text for a more detailed
description and for information on the adopted values for the cross sections.}
\label{tab:xsec}
\end{center}
\end{table}

Focusing first on the signal, we generate, for each scenario and each sgluon
mass (see Table \ref{tab:params} and Table \ref{tab:sigma}), three event samples
according to the top
multiplicity of the final state, the latter being taken equal to two, three
and four top quarks, respectively. Our event sample normalization
includes the NLO $K$-factor presented in Table \ref{tab:sigma}. 
Turning to the background, we are planning to 
require, in our analysis, the presence of at least one isolated charged lepton.
We therefore only simulate the Standard Model background
processes presented in Table \ref{tab:xsec}, which also contains, in addition to
the generation parameters above-mentioned, the cross sections employed for the
normalization of the samples and the associated numbers of generated
events. 

We first address the simulation of weak gauge boson production in association
with jets. We have merged event samples containing up to four additional hard
jets and the gauge bosons are enforced to decay leptonically. Parton-level
events having also been allowed to contain tau leptons, we make use of
the {\sc Tauola} program \cite{Davidson:2010rw} in order
to handle their decays. We have
reweighted the events according to the next-to-next-to-leading order (NNLO)
cross sections as computed by the {\sc Fewz} package
\cite{Melnikov:2006kv,Gavin:2012kw,Gavin:2010az} with the recent set of parton
densities CT10 provided by the CTEQ collaboration \cite{Lai:2010vv}.
Virtual photon contributions are included where relevant, together with a
parton-level selection based on the dilepton invariant mass $m_{\ell\ell} > 50$
GeV.

We now turn to background events related to top quark production. We generate
two distinct $t \bar t$ samples, one associated to the
semileptonic decay of the $t\bar t$ pair and one related to its dileptonic
decay. Our merging procedure includes matrix elements containing up
to two additional jets and we have reweighted all the events according to
the production cross section at the NLO accuracy, including
genuine NNLO contributions, as predicted by the {\sc Hathor} program 
\cite{Aliev:2010zk,Baernreuther:2012ws}. 
Single top event generation has been split into the generation of three
different inclusive samples, following the usual parton-level distinction
between $s$-channel diagrams where the top quark is produced in
association with a $b$ quark, $t$-channel diagrams where the top quark is
produced in association with a light jet, and $tW$ diagrams describing the
associated production of a top quark and a $W$-boson. In order to maintain this
distinction non-ambiguous, the MLM merging procedure has not been applied. This
avoids a possible double
counting over the three channels since specific diagrams with extra radiation
could in principle belong to several of the categories, even though the
kinematical regimes are different. The events are
then reweighted according to NLO cross sections
including genuine NNLO contributions \cite{Kidonakis:2010tc,
Kidonakis:2010ux, Kidonakis:2011wy, Kidonakis:2012db}.

Concerning diboson production, we have
performed a merging of matrix elements including up to two additional hard jets
and have normalized the cross sections to the NLO accuracy as provided by
the {\sc Mcfm} package \cite{Campbell:1999ah, Campbell:2011bn}. In our setup, we 
have included virtual photon contributions where relevant and consequently
imposed a selection on
the dilepton invariant-mass of $m_{\ell\ell} > 50$ GeV.

Rare Standard Model processes where a top-antitop pair is produced in
association with one or two gauge bosons are also considered. We generate
inclusive event samples for the $t\bar t W$, $t\bar t Z$ and $t\bar t WW$
processes and perform MLM-merging of matrix elements containing up to two extra
jets. We normalize the samples to the NLO cross sections as predicted by
{\sc Mcfm} \cite{Campbell:2012dh} for the $t\bar t W$ and $t \bar t Z$ 
processes and to the LO results as returned by {\sc MadGraph} for the $t\bar t
WW$ process. Finally, four-top production is also simulated but
no merging is performed since the lowest order matrix-element
already contains many hard jets. This event sample is normalized to the LO
accuracy.


\section{Probing sgluons via multitop events at the LHC}\label{sec:analysis}
In this paper, we wish to estimate the LHC sensitivity to the search for
multitop events originating from sgluon decays. According to the two classes of
scenarios under consideration (see Table \ref{tab:params} and Table
\ref{tab:sigma}), sgluon pair production and decay lead to three topologies
comprised of two top quarks and two light jets ($tjtj$), three top quarks and one
light jet ($tjtt$) and four top quarks ($tttt$), the symbol $t$ denoting
equivalently a top and an antitop quark and the symbol $j$ a light jet or
$b$-jet. The first two signatures are only considered assuming scenarios of class
I, while four-top final states are produced in the context of both classes of
scenarios. The 
corresponding final states are thus characterized by a large number of hard jets
(between four and twelve) with an important heavy-flavor content arising from
top decays. The large branching ratio associated with top quark hadronic
decays would naively encourage us to search for sgluon pairs in fully hadronic
final states. This analysis would however suffer from an overwhelming multijet
background whose a correct estimation requires data-driven methods. To
optimize the sensitivity of our search, we therefore restrict ourselves to
leptonic final states. We design two analyses, one dedicated to events
containing exactly one single lepton and another one to events with at least two
leptons.

\subsection{Event selection for a multilepton plus jets
signature}\label{sec:multilepsel}
Events are preselected with the requirement that they contain exactly two (for
the $tjtj$ topology) or at least two charged leptons (for the $tjtt$ and $tttt$
topologies) with a transverse momentum $p_T^\ell > 20$ GeV. The invariant mass
of the pair comprised of the two leading leptons $m_{\ell \ell}$ is also requested to
be larger than 50 GeV in order to be compatible with the parton-level selection
criterion of Section \ref{sec:mc}. At this stage, the total Standard Model
background contains a large part of Drell-Yan lepton pair events (98.7\% and 98.2\%
for the $tjtj$ and $tjtt$/$tttt$ topologies, respectively). To reduce this
background, a selection on the missing transverse energy $\slashed{E}_T$, defined
as 
\be
  \slashed{E}_T = \bigg|\bigg| \sum_{\text{visible particles}} \vec p_T \bigg|
    \bigg| \ , 
\ee 
is applied. By requiring $\slashed{E}_T > 40$ GeV, we take advantage of the fact
that Drell-Yan events are characterized by a lack of missing energy whereas
the neutrinos arising from leptonic top decays ensure signal events to contain a
sensible quantity of missing energy.

\begin{figure}
  \begin{center}
  \includegraphics[width=0.49\columnwidth]{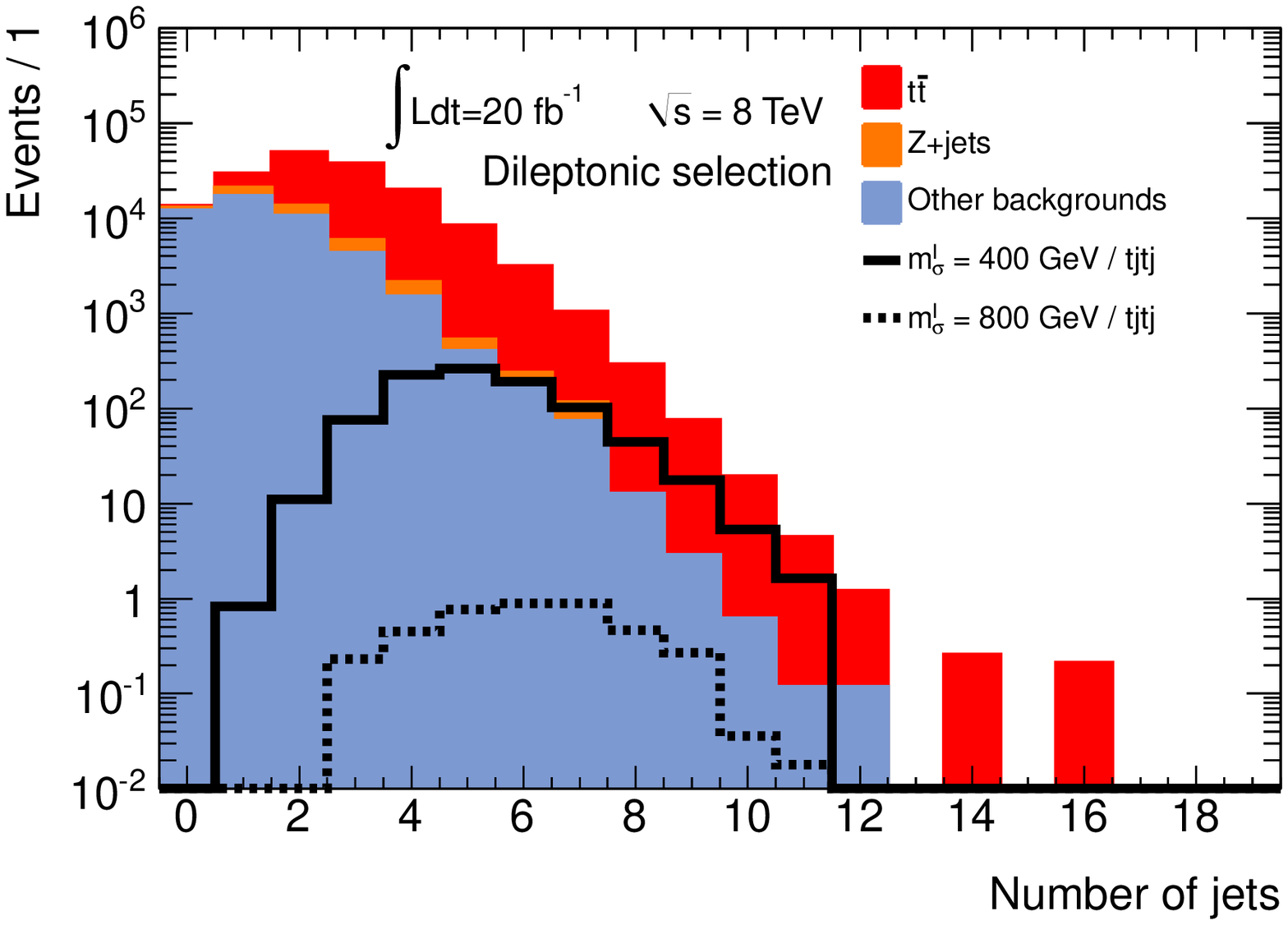}
  \includegraphics[width=0.49\columnwidth]{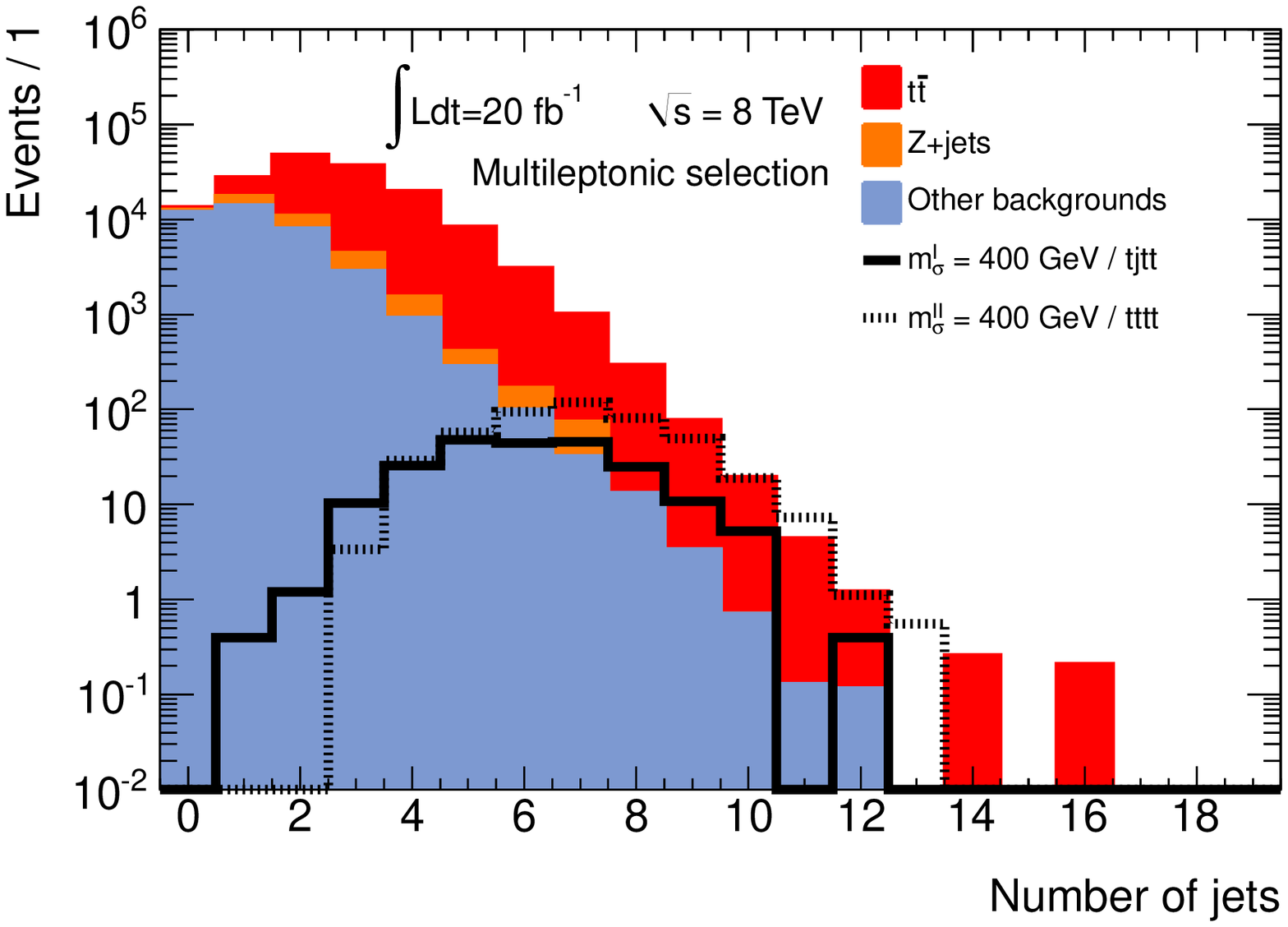}
  \caption{Jet multiplicity distribution after selecting events with exactly
    (left panel) or at least (right panel) two leptons, missing transverse
energy $\slashed{E}_T > 40$ GeV and a dilepton invariant-mass $m_{\ell\ell} > 50$
GeV. Contributions arising from top-antitop (red) and Drell-Yan
production (orange) are factorized from the rest of the background (blue)
and signal distributions for the $tjtj$ signature (left panel) and for both the
$tjtt$ and $tttt$ channels (right panel) are indicated by plain and dashed
curves.}
    \label{fig:n_jets_2l}
  \end{center}
\end{figure}

Jets present in sgluon-pair events originate mainly from the hadronization of
$b$-quarks and light quarks issued from top decays. In contrast, the hadronic
activity of background events consists dominantly of initial-state radiation that
leads to a lower jet multiplicity (see Figure \ref{fig:n_jets_2l}). To maintain
simultaneously a good sensitivity
to the sgluon signal while discarding a substantial part of the background, the
presence of at least three, four and five jets with $p_T^j > 25$ GeV is demanded
for the $tjtj$, $tjtt$ and $tttt$ final states, respectively. In addition, we
benefit from the presence of jets arising from the fragmentation of long-lived
$b$-quarks in sgluon-induced events, requiring a minimum number of one, two and
three $b$-tagged jets for the $tjtj$, $tjtt$ and $tttt$ topologies,
respectively.

\begin{table}[t] 
\begin{center}
\begin{tabular}{ | c || c c c |}
\hline 
  \multirow{2}{*}{Selections}& \multicolumn{3}{c|}{ $tjtj$ channel}  \\ 
   & $m_\sigma^I = 400$ GeV & $m_\sigma^I = 800$ GeV &Backgrounds \\ 
\hline 
  $N_\ell = 2$ with $p^\ell_T \geq 20$ GeV  & 	$(1.26 \pm 0.02) \!\cdot\! 10^3$& 	$4.86 \pm 0.30$ & $(1.721 \pm 0.002) \!\cdot\! 10^7$ \\ 
  $m_{\ell\ell} \geq 50$ GeV  & 		$(1.15 \pm 0.02)\!\cdot\!10^3$ 	&     	$4.49 \pm 0.28$ & $(1.716 \pm 0.002) \!\cdot\! 10^7$ \\ 
  $\slashed{E}_T \geq 40$ GeV & 		$(9.38 \pm 0.20)\!\cdot\!10^2$ 	& 	$4.04 \pm 0.27$ & $(1.549 \pm 0.004) \!\cdot\! 10^5$\\ 
  $N_j\geq 3$ with $p^j_T \geq 25$ GeV & 	$(9.18 \pm 0.19) \!\cdot\! 10^2$&    	$4.04 \pm 0.27$ & $(5.693 \pm 0.020) \!\cdot\! 10^4$ \\ 
  $N_b\geq 1$ & 				$(6.05 \pm 0.16) \!\cdot\! 10^2$& 	$2.80 \pm 0.22$ & $(4.089 \pm  0.011) \!\cdot\! 10^4$ \\ 
  Same sign dilepton & 				$(2.81 \pm 0.11) \!\cdot\! 10^2$& 	$1.06 \pm 0.14$ & $(4.191 \pm 0.035) \!\cdot\! 10^3$ \\ 
\hline 
\end{tabular}\vspace{.25cm} 

\begin{tabular}{ | c || c c c |}
\hline 
  \multirow{2}{*}{Selections}& \multicolumn{3}{c|}{ $tjtt$ channel}  \\ 
   & $m_\sigma^I = 400$ GeV & $m_\sigma^I = 800$ GeV &Backgrounds \\ 
\hline 
  $N_\ell \geq 2$ with $p^\ell_T \geq 20$ GeV & $(2.89 \pm 0.11) \!\cdot\! 10^2$ & $4.71 \pm 0.17$ & $(1.722 \pm 0.002) \!\cdot\! 10^7$ \\ 
  $m_{\ell\ell} \geq 50$ GeV & 			$(2.63 \pm 0.10) \!\cdot\! 10^2$ & $4.44 \pm 0.17$ & $(1.717 \pm 0.002) \!\cdot\! 10^7$ \\ 
  $\slashed{E}_T \geq 40$ GeV & 		$(2.17 \pm 0.09) \!\cdot\! 10^2$ & $4.12 \pm 0.16$ & $(1.598 \pm 0.004) \!\cdot\! 10^5$ \\ 
  $N_j\geq 4$ with $p^j_T \geq 25$ GeV & 	$(1.97 \pm 0.09) \!\cdot\! 10^2$ & $4.03 \pm 0.16$ & $(2.375 \pm 0.012) \!\cdot\! 10^4$ \\ 
  $N_b\geq 2$ & 				$83 \pm 6$ 			 & $1.89 \pm 0.11$ & $(5.950 \pm 0.040) \!\cdot\!  10^3$ \\ 
  Same sign dilepton & 				$36 \pm 4 $ 			 & $0.77\pm 0.07$ & $(2.860\pm 0.080)  \!\cdot\! 10^2$ \\ 
\hline 
\end{tabular} \vspace{.25cm}
 
\begin{tabular}{ | c || c c c |}
\hline 
  \multirow{2}{*}{Selections}& \multicolumn{3}{c|}{ $tttt$ channel}  \\ 
   & $m_\sigma^I = 400$ GeV & $m_\sigma^{II} = 800$ GeV &Backgrounds \\ 
\hline 
  $N_\ell \geq 2$ with $p^\ell_T \geq 20$ GeV 	& \phantom{$ \ \, $} 11.33 $\pm$ 0.33\phantom{$\ \, $} 	& 7.90 $\pm$ 0.24 & $(1.722 \pm 0.002) \!\cdot\! 10^7$ \\ 
  $m_{\ell\ell} \geq 50$ GeV 			& 10.42 $\pm$ 0.32 			& 7.56 $\pm$ 0.22 & $(1.717 \pm   0.002) \!\cdot\! 10^7$ \\ 
  $\slashed{E}_T \geq 40$ GeV			&8.78 $\pm$ 0.29 			& 7.03 $\pm$ 0.21 & $(1.598 \pm 0.004) \!\cdot\! 10^5$ \\ 
  $N_j\geq 5$ with $p^j_T \geq 25$ GeV 		& 7.50 $\pm$ 0.27			& 6.60 $\pm$ 0.20 &  $(8.11 \pm 0.06) \!\cdot\!10^3$ \\ 
  $N_b\geq 3$  					&1.61 $\pm$ 0.13 			& 1.93 $\pm$ 0.11 & $(1.88 \pm 0.06) \!\cdot\! 10^2$\\
  Same sign dilepton  				&0.69 $\pm$ 0.08 			& 0.82 $\pm$ 0.07 &10.3 $\pm$ 1.5 \\ 
\hline 
\end{tabular}
\caption{Flow-charts related to the selection strategy for the $tjtj$ (upper
panel), $tjtt$ (middle panel) and $tttt$ (lower panel) topologies in the case of
a multilepton analysis. We present
the remaining number of events, together with their associated statistical
uncertainties, after each of the selection criteria in the context of the LHC
collider running at a center-of-mass energy of $\sqrt{s}=8$ TeV and 
for an integrated luminosity of 20 fb$^{-1}$. For the $tjtj$ and
$tjtt$ channels, signal events are given for scenarios of class I and for a
sgluon mass of $m_\sigma^I=400$ (800) GeV in the second (third) column of the
tables. Concerning the $tttt$ channel, we present in the second (third) column
of the table the evolution of the number of signal events for a
scenario of class I (II) with a sgluon mass of $m_\sigma^I = 400$ GeV
($m_\sigma^{II} = 800$ GeV) after applying each
of the selections. The associated numbers of background events are shown in the fourth
column of the tables.
For a detailed description of each of the selections, we refer to the text.} 
\label{tab:cuts_DILEPTON} 
\end{center}
\end{table} 

After applying the above-mentioned selections to the preselected dileptonic events, the
selection efficiencies for the signal are found to range from 15\% to 50\% for a
sgluon mass of $m_\sigma = 400$ GeV and from 25\% to 60\% for $m_\sigma = 800$ GeV
for the three  search channels (derived from Table
\ref{tab:cuts_DILEPTON}). The expected Standard Model
background, which has been divided by a factor of about 400, 3000 and 100000 for
the $tjtj$, $tjtt$ and $tttt$ search strategies, respectively, is now largely
comprised of
Drell-Yan and top-antitop (plus possibly one or two additional gauge bosons)
events. To further reduce it, we apply a
specific selection on the fraction of dileptonic events and only retain those
where the leptons have the same electric charge. The signal efficiency of such a
criterion is of about 50\% while only 10\% and 20\% of the background
events survive in the context of the $tjtj$ and $tjtt$/$tttt$ topologies. 

The numbers of events selected at each step of the analysis are indicated in
Table \ref{tab:cuts_DILEPTON} for two representative benchmark scenarios with
sgluon masses of 400 GeV and 800 GeV, as well as for the background. It turns
out that after all selection criteria, the background is dominated by top-antitop events for
the $tjtj$ and
$tjtt$ search channels and by events related to the associated production of a
top-antitop pair with one or two additional gauge bosons for the $tttt$
signature.

In our simulation setup, the multijet background, jets faking leptons and charge
misidentification have 
not been accounted for. However, on the basis of the analysis of Ref.\ 
\cite{ATLAS-CONF-2012-130} where same sign dilepton events are investigated
after selection criteria similar to those applied in this paper, these sources
of background have been found to contribute in a significant way as the
associated number of surviving events after applying all the selections is ten
times larger. We
therefore adopt a conservative approach and derive, in Section
\ref{sec:resu}, two limits on sgluon-induced new physics. The first one is
extracted after omitting the non-simulated backgrounds and the second one is
obtained after multiplying the number of background events by a factor of ten.

\begin{figure}[t!]
  \begin{center}
    \includegraphics[width=0.49\columnwidth]{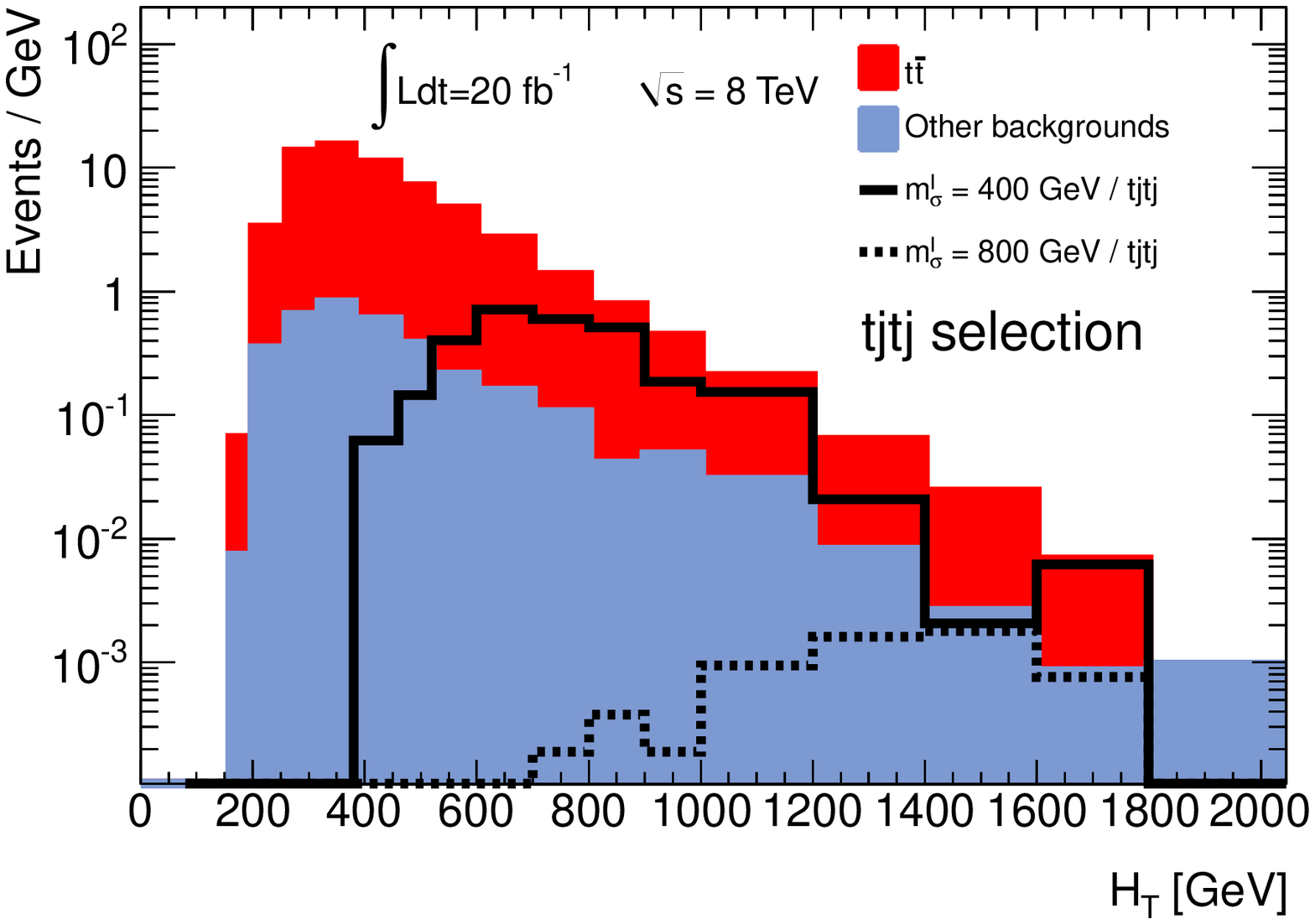}
    \includegraphics[width=0.49\columnwidth]{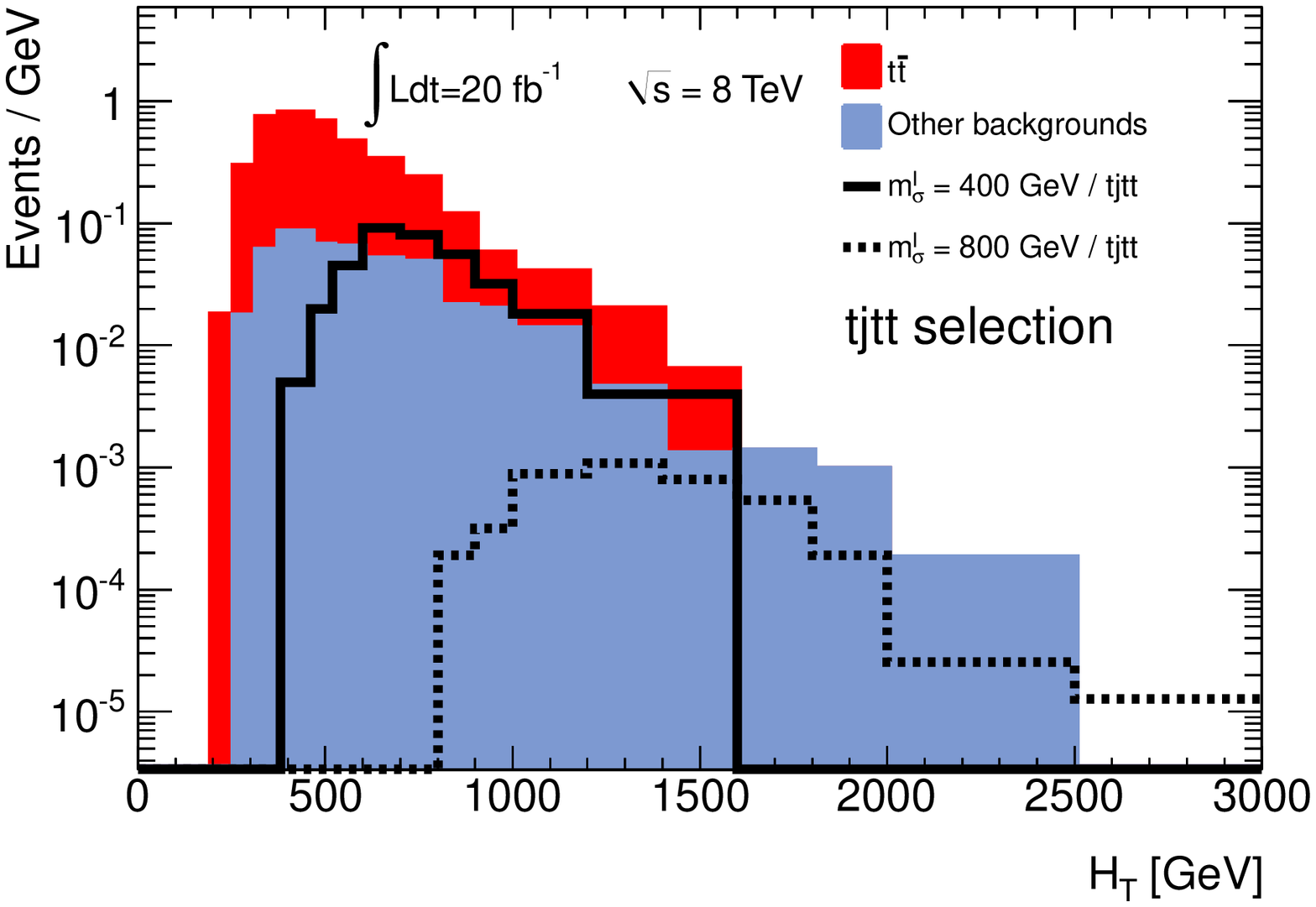}
    \caption{Distribution of the $H_T$ variable defined in Eq.\
  \eqref{eq:htdef} after applying the selection strategy associated with the
  multilepton analysis presented in the text for the $tjtj$ (left) and $tjtt$
  (right) topologies. We distinguish the dominant source of background 
  associated with the production of $t \bar t$ pairs together with jets (red) from
  the other contributions (blue). We superimpose the corresponding curves for two
  signal scenarios of class I with respective sgluon
  masses   of 400 GeV (plain black curve) and 800 GeV (dashed black curve).}
    \label{fig:HT_dil}
  \end{center}
\end{figure}

We then consider the $H_T$ variable defined by
\be\label{eq:htdef}
   H_T = \sum_{\text{jets, leptons, missing energy}} \big|\big| \vec p_T \big| \big| \
   ,
\ee
since signal events are expected to contain more jets and leptons than
background events. Omitting the $tttt$ channel as its
statistical significance is very poor (see Table
\ref{tab:cuts_DILEPTON} for illustrative benchmark scenarios), we present
$H_T$ distributions for the 
$tjtj$ and $tjtt$ topologies on the left and right panels of Figure
\ref{fig:HT_dil}, respectively. On both panels, we show curves associated with
signal scenarios of class I where the sgluon mass is set to 400 GeV and 800
GeV. The distributions present a steep rise once the production
threshold is reached, followed by a large peak centered around twice the sgluon mass. 
We then compare these shapes to the corresponding background distributions, after
factorizing out the dominant $t\bar t$ contribution (in red) from the rest of
the background events (in blue). This suggests to probe the LHC sensitivity to
the presence of sgluon fields coupling
dominantly to top quarks by means of a careful investigation of the shape of the
entire $H_T$ distributions, keeping all events, rather than requiring
(unefficient) selection on this variable. Limits at the 95\% confidence level
are extracted in this way in  Section \ref{sec:resu}.

\subsection{Event selection for a single lepton plus jets signature}
\label{sec:monolepsel}
Final states containing exactly one lepton are expected to be copiously produced
from sgluon pair production and decay at the LHC as they correspond to a fraction 
of events equal to 36\% and $\sim 41\%$ in the context of the $tjtj$ and
$tjtt$/$tttt$ topologies, respectively. We preselect events by requiring exactly
one single lepton with a transverse momentum $p_T^\ell > 25$ GeV. Consequently, the
Standard Model background is expected to be dominated at $92\%$ by events
associated with the production of a $W$-boson in association with jets. The
expected number of sgluon events ranges from 34.9 (24.6)
to 10700 (48.4) for a representative scenario with a sgluon mass of 400 GeV (800
GeV), as shown in Table \ref{tab:cuts_MONOLEPTON}.

\begin{table}[t] 
\begin{center}
\begin{tabular}{ | c || c c c |}
\hline 
  \multirow{2}{*}{Selections}& \multicolumn{3}{c|}{ $tjtj$ channel}  \\ 
   & $m_\sigma^I = 400$ GeV & $m_\sigma^I = 800$ GeV &Backgrounds \\ 
\hline 
  $N_\ell = 1$ with $p^\ell_T \geq 25$ GeV & 	$(1.06 \pm 0.01) \!\cdot\! 10^4$ &  $ 45.7 \pm 0.9 $ & $(2.376 \pm 0.003) \!\cdot\!10^8$ \\ 
  $\slashed{E}_T \geq 40$ GeV  & 		$(7.65 \pm 0.06) \!\cdot\! 10^3$ &  $ 37.9 \pm 0.8 $ & $(6.836 \pm 0.002) \!\cdot\! 10^7$ \\ 
  $M_T^W \geq 25$ GeV & 			$(6.43 \pm 0.05) \!\cdot\! 10^3$ &  $ 30.7 \pm 0.7 $ & $(6.722 \pm 0.002) \!\cdot\! 10^7$ \\ 
  $N_j \geq 6$ with$p_T^j \geq 25$ GeV & 	$(3.88 \pm 0.04) \!\cdot\! 10^3$ &  $ 24.9 \pm 0.7 $ & $(8.634 \pm 0.024) \!\cdot\! 10^4$ \\ 
  $N_b\geq 1$ & 				$(2.91 \pm 0.04)\!\cdot\! 10^3$  &  $ 19.3 \pm 0.6 $ & $(6.407 \pm 0.014) \!\cdot\! 10^4$ \\ 
\hline 
\end{tabular} \vspace{.25cm} 

\begin{tabular}{ | c || c c c |}
\hline 
  \multirow{2}{*}{Selections}& \multicolumn{3}{c|}{ $tjtt$ channel}  \\ 
   & $m_\sigma^I = 400$ GeV & $m_\sigma^I = 800$ GeV &Backgrounds \\ 
\hline 
  $N_\ell = 1$ with $p^\ell_T \geq 25$ GeV 	& $(1.21 \pm 0.22) \!\cdot\! 10^3$	&$21.3 \pm 0.4$ & $(2.376 \pm 0.001) \!\cdot\!10^8$ \\ 
  $\slashed{E}_T \geq 40$ GeV 			& $(8.81 \pm 0.19) \!\cdot\! 10^2$ 	&$18.1 \pm 0.3$ & $(6.836 \pm 0.002) \!\cdot\!10^7$ \\ 
  $M_T^W \geq 25$ GeV  				& $(7.66 \pm 0.18)\!\cdot\! 10^2$ 	&$15.4 \pm 0.3$ & $(6.722 \pm 0.002)\!\cdot\! 10^7$ \\ 
  $N_j \geq 7$ with $p_T^j \geq 25$ GeV 	& $(4.05 \pm 0.13)\!\cdot\! 10^2$ 	&$11.08 \pm 0.3$ & $(2.613 \pm 0.012)\!\cdot\! 10^4$ \\ 
  $N_b\geq 2$ 					& $(1.99 \pm 0.09) \!\cdot\! 10^2$ 	&$5.99 \pm 0.2$ & $(9.330\pm 0.050) \!\cdot\! 10^3$ \\ 
\hline 
\end{tabular} \vspace{.25cm} 

\begin{tabular}{ | c || c c c |}
\hline 
  \multirow{2}{*}{Selections}& \multicolumn{3}{c|}{ $tttt$ channel}  \\ 
   & $m^I_\sigma \ = \ 400$ GeV & $m^{II}_\sigma = 800$ GeV &Backgrounds \\ 
\hline 
  $N_\ell = 1$ with  $p^\ell_T \geq 25$ GeV 	& $34.6 \pm 0.6$ 	& $23.2 \pm 0.4$ & $(2.376 \pm 0.001)\!\cdot\! 10^8$ \\ 
  $\slashed{E}_T \geq 40$ GeV 			& $27.3 \pm 0.5$ 	& $20.2 \pm 0.4$ & $(6.836 \pm 0.002) \!\cdot\! 10^7$ \\ 
  $M_T^W \geq 25$ GeV 				& $23.6 \pm 0.5$ 	& $17.1 \pm 0.3$ & $(6.722 \pm 0.002) \!\cdot\! 10^7$ \\ 
  $N_j \geq 8$ with $p_T^j \geq 25$ GeV 	& $10.8 \pm 0.3$ 	& $12.3 \pm 0.3$ &$(7.020 \pm 0.060) \!\cdot\! 10^3$ \\ 
  $N_b\geq 2$ 					& $7.21 \pm 0.27$	& $8.47 \pm 0.23$ 	& $(2.658 \pm 0.026)  \!\cdot\! 10^3$ \\ 
\hline 
\end{tabular} 
\caption{Same as Table \ref{tab:cuts_DILEPTON} but in the context of a single
lepton final state. 
For a detailed description of each of the selection criteria, we refer to the text.} 
\label{tab:cuts_MONOLEPTON} 
\end{center}
\end{table}

The multijet background has not been taken into account up to now. However,
data-driven methods used to estimate its shape and normalization tend to show
that it may have to be considered in our analysis \cite{Aad:2012wm}. In order to
realistically and reliably reject this
background, we impose, inspired by the experimental analysis of Ref.\
\cite{Aad:2012wm}, the missing transverse energy of the events to be larger
than $\slashed{E}_T > 40$ GeV and the reconstructed $W$-boson transverse mass,
defined as
\be
  M_T^W =  \sqrt{ 2 p_T^\ell \slashed{E}_T \Big[ 1 - \cos \Delta
    \phi_{\ell,\slashed{E}_T}\Big] } \ , 
  \label{Eq:TransverseMasse}
\ee
to be larger than 25 GeV. In this equation, the quantity $\Delta\phi_{\ell,
\slashed{E}_T}$ stands for the angular distance, in the azimutal direction with
respect to the beam, between the lepton and the missing energy.

\begin{figure}
  \begin{center}
    \includegraphics[width=0.6\columnwidth]{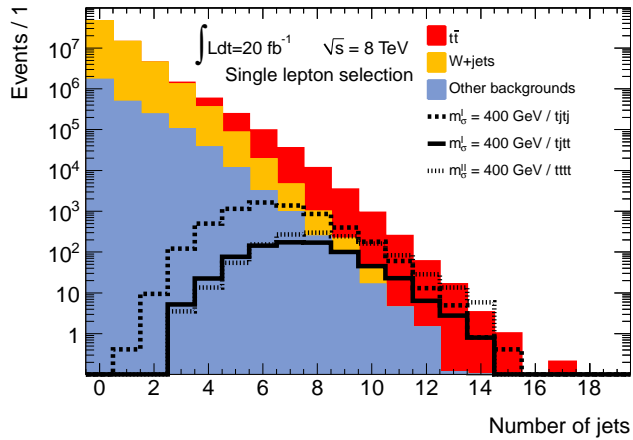}
    \caption{Jet multiplicity distribution after selecting events with exactly 
    one lepton, missing transverse energy $\slashed{E}_T > 40$ GeV and a
$W$-boson transverse mass $M_T^W > 25$ GeV. We distinguish the $t\bar t$ (red)
and $W$-boson plus jets (orange) contributions from the rest of the background
(blue) and present signal distributions for a sgluon scenarios of class I with
$m_\sigma = 400$ GeV in the $tjtj$ (plain), $tjtt$ (strong dashed) and $tttt$
(light dashed) channels.}
    \label{fig:n_jets_1l}
  \end{center}
\end{figure}

As for the multilepton analysis, a higher multiplicity of hard jets is expected in
signal events. They indeed arise from the hadronization of final state quarks,
in contrast to the $W$-boson plus jets background where jets originate
predominantly from initial-state radiation (see Figure \ref{fig:n_jets_1l}). 
Events are therefore selected with the requirement that they contain at least
six, seven and eight jets with a transverse-momentum $p_T^j > 25$ GeV for the
$tjtj$, $tjtt$ and $tttt$ topologies, respectively. Moreover, sgluon events are
expected to include a higher number of heavy-flavored jets. The minimal number
of $b$-tagged jets is therefore
required to be larger that one and two for the $tjtj$ and $tjtt$/$tttt$ search
channels, respectively. At this stage, the expected Standard Model background is
composed mainly of $t\bar t$ events, where the top-antitop pair is possibly
produced in association with one or several gauge bosons.

The number of events surviving to each of the selection criteria is
presented in Table
\ref{tab:cuts_MONOLEPTON} for two representative signal scenarios with $m_\sigma
= 400$ GeV and 800 GeV as well as for the background. After all selections, we expect
a number of signal events ranging from 7.28 (8.96) to 2940 (20.4) for a sgluon
mass of 400 (800) GeV according to the search channel under consideration. In
contrast, the Standard Model predicts a background of 64070, 9330 and 2658
events for the $tjtj$, $tjtt$ and $tttt$ topologies.

\begin{figure}
  \begin{center}
    \includegraphics[width=0.49\columnwidth]{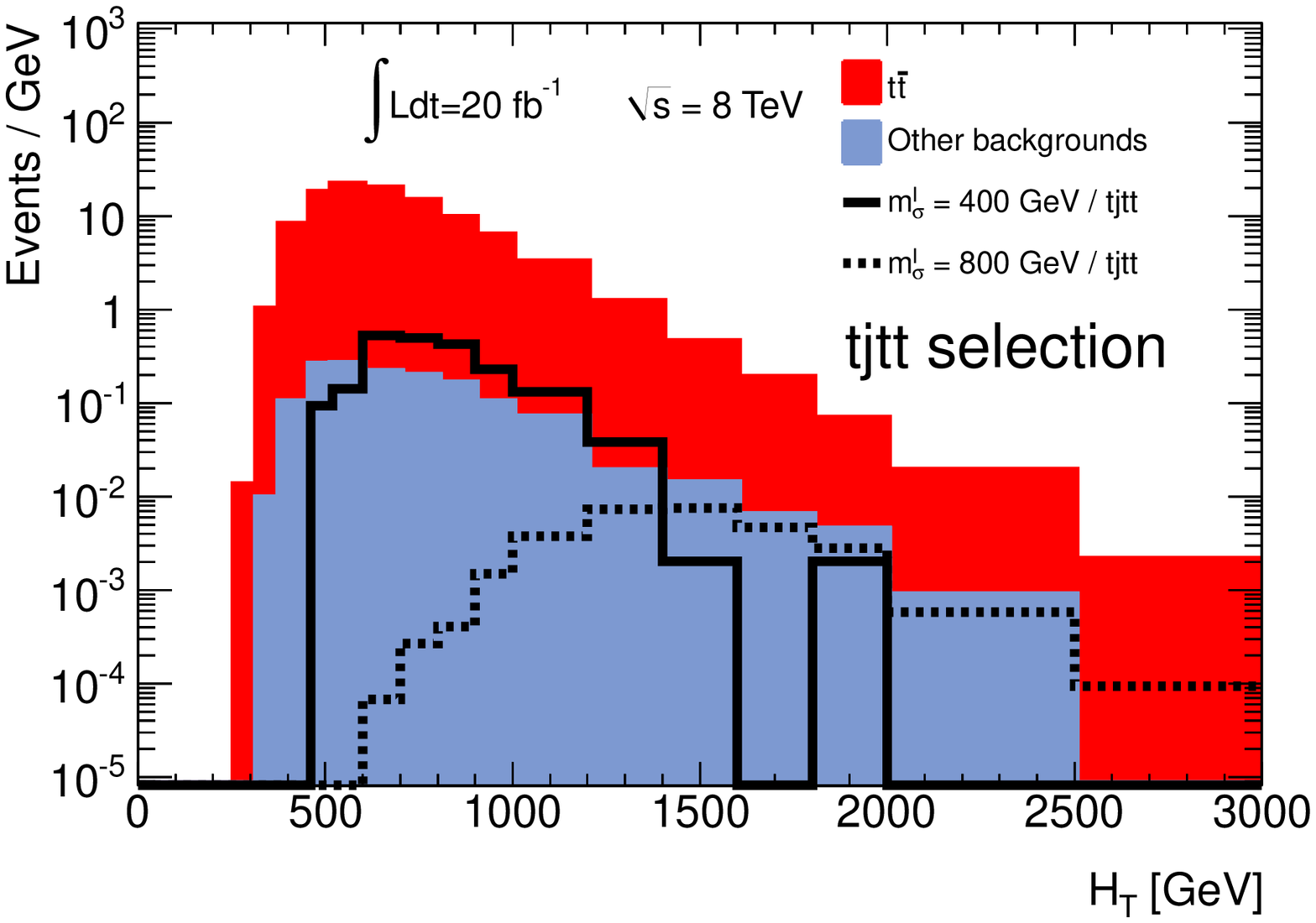}
    \includegraphics[width=0.49\columnwidth]{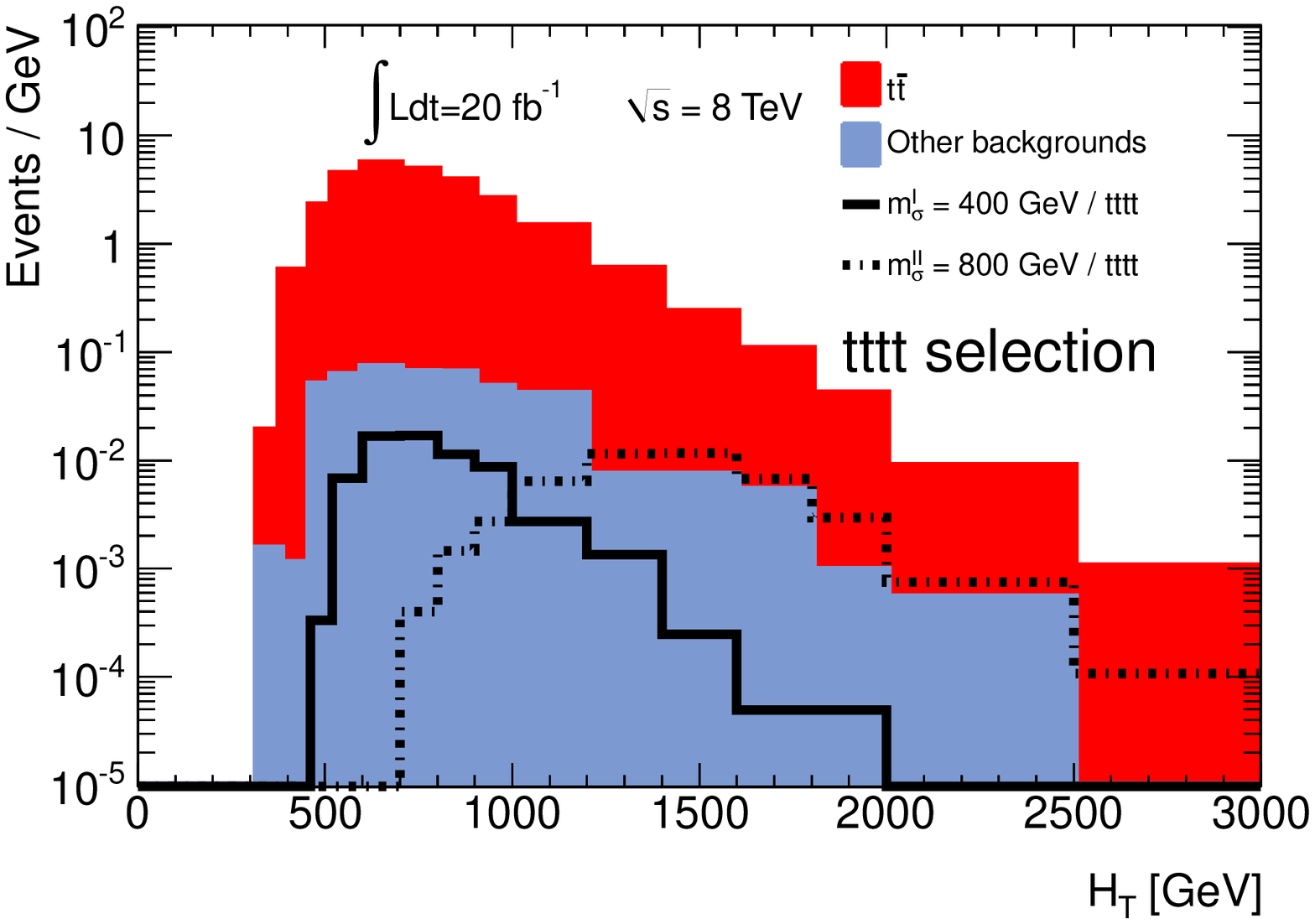}
    \caption{Distribution of the $H_T$ variable defined in Eq.\
  \eqref{eq:htdef} after the selection strategy associated to a single lepton
  analysis as presented in the text. We distinguish the background events
  associated to the production of $t \bar t$ pairs together with jets (red) from
  the other contributions (blue). For the $tjtt$ channel (left panel), we
  superimpose the corresponding curves for two
  signal scenarios of class I with respective sgluon
  masses of 400 GeV (plain black curve) and 800 GeV (dashed black curve). In the
  case of the $tttt$ search-channel (right panel), we rather consider a scenario
  of class I with a sgluon mass of 400 GeV (plain black curve) and a scenario of
  class II with a sgluon mass of 800 GeV (dashed black curve).}
    \label{fig:HT_1l}
  \end{center}
\end{figure}

Due to the larger activity expected in signal events compared to
background events, we consider, as in Section \ref{sec:multilepsel}, the
$H_T$ variable as a discriminant between signal and background.
Following the definition of Eq.\ \eqref{eq:htdef},
we present, in Figure \ref{fig:HT_1l}, the $H_T$ distributions after
following the $tjtt$ (left panel) and $tttt$ (right panel) single lepton
selection strategy introduced above. We
show background distributions after distinguishing events associated with
top-antitop production in association with jets (red) from the rest of the
background (blue). The resulting behavior underlines a steep rise once the
top-antitop production threshold is reached, both in the $tjtt$ and $tttt$
channels, followed by a peak around
$H_T\sim 500$ GeV and a smooth fall with increasing energy.
In comparison, signal distributions are associated with a clear peaky behavior
centered around an $H_T$ value depending on the sgluon mass ($\sim 1.5
m_\sigma$) and the tail of the distributions does not extend to a very large
hadronic energy, in contrast to the Standard Model background expectation.
Instead of requiring a selection criterion based on the $H_T$
variable, we then employ the shape of the associated distributions as inputs
to extract the LHC sgluon mass reach in Section \ref{sec:analysis}.

In the case of a $tjtj$ signature with exactly one final state lepton, the
reconstruction of the sgluon mass is possible if the missing energy is assumed
to originate only from the neutrino issued from the $W$-boson decay.
Assigning the labeling of the six jets according to the process
\be
  p p \to \sigma \sigma \to (t j_5) (t j_6) \to (j_1 j_2 j_3 j_5) (j_4 \ell \nu
   j_6) \ ,
\label{eq:chi2pattern}\ee
all possible permutations of the six jets are performed
and the one minimizing a $\chi^2$-variable defined by 
\be\bsp
  \chi^2  =&\ 
  \left[ \frac{m_{j_1j_2}-m^{(r)}_W}{\sigma^{(r)}_W} \right]^{2} + \left[
    \frac{\big(m_{j_1j_2j_3}-m_{j_1j_2}\big) - m^{(r)}_{tW}}{\sigma^{(r)}_{tW}}
    \right]^{2} +
    \left[ \frac{m_{\ell\nu j_4}-m^{(r)}_{t\ell}}{\sigma^{(r)}_{t\ell}}
    \right]^{2} + \\
  &\  \left[\frac{(m_{\ell\nu j_4,j_6}-m_{\ell\nu
    j_4})-(m_{j_1j_2j_3,j_5}-m_{j_1j_2j_3})}
    {\sigma_{\sigma t}^{(r)} \big[ (m_{\ell\nu j_4,j_6}-m_{\ell\nu
    j_4})+(m_{j_1j_2j_3,j_5}-m_{j_1j_2j_3}) \big] } \right]^2
\esp\label{eq:chi2} \ee
is retained as the true configuration of a given event. 
The analytical expression of this $\chi^2$-variable exactly mimics Eq.\
\eqref{eq:chi2pattern}.
The three jets $j_1$, $j_2$ and $j_3$ are the ones issued from
the hadronically decaying top quark, $j_3$ being a 
$b$-jet\footnote{We are however not making use of jet flavor information in our 
kinematical fit.}, and this information is encompassed into the first two terms
of the $\chi^2$ variable.  The invariant mass of the 
two light jets $j_1$ and $j_2$, denoted by $m_{j_1 j_2}$, is asked to be
compatible with the 
$W$-boson mass and the three-jet invariant mass $m_{j_1j_2j_3}$ is required to
be compatible with the top mass. However, 
in order not to introduce a correlation between the first two terms of the
$\chi^2$ variable, we
subtract from the reconstructed top mass $m_{j_1j_2j_3}$ the reconstructed dijet
invariant-mass $m_{j_1j_2}$. The values of the fit parameters are
taken as $m^{(r)}_W =
80.7$ GeV, $\sigma^{(r)}_W = 8.9$ GeV,  $m^{(r)}_{tW} = 90.8$~GeV and
$\sigma^{(r)}_{tW} = 10.5$ GeV. These numerical values have been extracted from
a fit based on the Monte Carlo truth, which ensures that each reconstructed
object is correctly assigned according to the configuration of Eq.\
\eqref{eq:chi2pattern}. In our simulation setup, it must be noted that the large
values of ${\cal O}(10\%)$ for the widths are mainly related to detector
resolution.

In the third term of Eq.\ \eqref{eq:chi2}, we focus on the leptonically decaying
top quark and ask the invariant mass $m_{\ell\nu j_4}$ to be compatible with the
top mass. Since
the neutrino four-momentum is entirely reconstructed from the assumption that
both the missing energy and the charged lepton are issued from the decay of 
a $W$-boson, there is no need to add an extra term in the definition of the
$\chi^2$, the information being already implicitly included in the
$m_{\ell\nu j_4}$ term. From the Monte Carlo truth, we have calculated a
reconstructed mass equal to $m^{(r)}_{tl} = 167.8$ GeV, together with a
width of $\sigma^{(r)}_{tl} = 19.1$ GeV. 

Finally, the definition of the $\chi^2$ variable of Eq.\ \eqref{eq:chi2} exploits
the fact that both 
the $tj_5$ and $tj_6$ pairs are issued from the decay of a sgluon field.
The corresponding invariant masses $m_{j_1j_2j_3, j_5}$ and $m_{\ell\nu
j_4,j_6}$ must hence be self-compatible, up to the detector
resolution. This is translated in the last term of Eq.\ \eqref{eq:chi2}, after
subtracting the respective reconstructed top masses to avoid possible
correlations among the different terms of the $\chi^2$. From the Monte Carlo
truth, we fix the related width to $\sigma_{\sigma t}^{(r)} = 0.098$ GeV.

\begin{figure}
  \begin{center}
    \includegraphics[width=0.65\columnwidth]{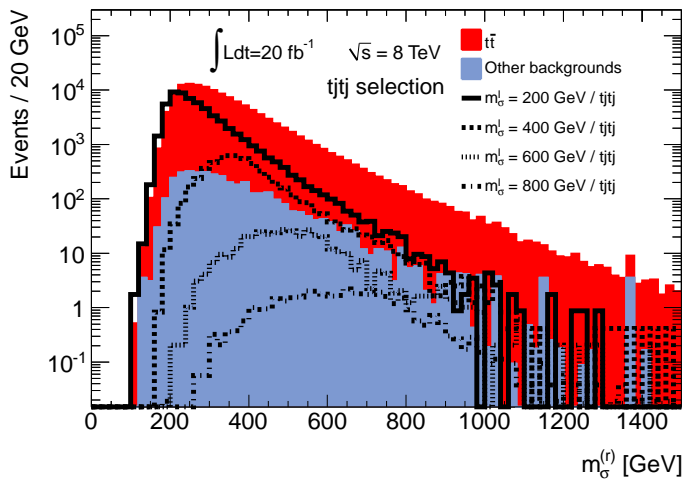}
    \caption{Reconstructed sgluon mass $m_\sigma^{(r)}$ for both
    the background and several signal scenarios in the context of the $tjtj$
    topology. Event selection is performed as in the upper panel
    of Table \ref{tab:cuts_MONOLEPTON} and we reconstruct each event
    according to the pattern given in Eq.\ \eqref{eq:chi2pattern} by means
    of the minimization of the $\chi^2$ variable of Eq.\ \eqref{eq:chi2}.
    Concerning the background distribution, we distinguish the dominant top-antitop
    contribution (red) from the rest of the background (blue) and superimpose
    signal curves for four scenarios of class I, with respective sgluon mass of 200
    GeV (plain), 400 GeV (strong dotted), 600 GeV (light dotted) and 800 GeV
    (dash-dotted).}
    \label{fig:mass_spectrum_tjtj}
  \end{center}
\end{figure}

Performing an event selection as described in the upper panel of Table
\ref{tab:cuts_MONOLEPTON} and minimizing the $\chi^2$ variable of Eq.\
\eqref{eq:chi2}, each event is reconstructed according to the
pattern given in Eq.\ \eqref{eq:chi2pattern}. This allows us to extract the 
sgluon mass $m_\sigma^{(r)}$. The resulting distributions are presented on Figure
\ref{fig:mass_spectrum_tjtj} for both the background and four signal scenarios
of class I with a sgluon mass parameter $m_\sigma$ taken equal to
200 GeV, 400 GeV, 600 GeV and 800 GeV, respectively. As expected, the background
distribution, where we
again distinguish the top-antitop pair contributions (in red) from the other
sources of background (in blue), presents a rising behavior once the
production threshold of a top-antitop pair is reached, quickly followed by 
a slow fall which extends to rather large values of the reconstructed
sgluon mass $m_\sigma^{(r)}$. In contrast, the signal distributions all show a
peak. For light sgluons ($m_\sigma=200$ GeV or 400
GeV), this peak is clearly centered around the true sgluon mass. In the case of
heavier sgluons
($m_\sigma=600$ GeV and 800 GeV), the
detector resolution renders the peak very wide and centered around a value
equal to about $70\%-80\%$ of the real sgluon mass. We then use, in the next
section, the reconstructed mass $m_\sigma^{(r)}$ to extract limits on the sgluon
mass reachable at the LHC.

\subsection{LHC sensitivity to a sgluon field dominantly coupling to top quarks}
\label{sec:resu}
For each of the final states considered in this paper, we combine the number of
expected signal and background events, including their corresponding statistical
uncertainties, to calculate upper limits on the signal cross section
at the 95\% confidence level. To this aim, we use the \texttt{CLs} technique
\cite{Read:2002hq} as
implemented in the {\sc MCLimit} software \cite{Web:McLimit}. We employ the
$H_T$ variable defined in Eq.\ \eqref{eq:htdef} to discriminate 
signal from background in the multilepton (for the $tjtj$, $tjtt$ and $tttt$
topologies) and single lepton (for the $tjtt$ and $tttt$ topologies) analyses
while the reconstructed sgluon mass is chosen in the case of a single lepton
analysis applied to a $tjtj$ final state\footnote{We have checked that
considering the reconstructed mass instead of the $H_T$ variable allows us to
improve the LHC sensitivity by about 15\%-20\% in the low mass region. This
choice is however irrelevant for the higher sgluon mass region.}.

\begin{figure}
  \begin{center}
    \includegraphics[width=0.56\columnwidth]{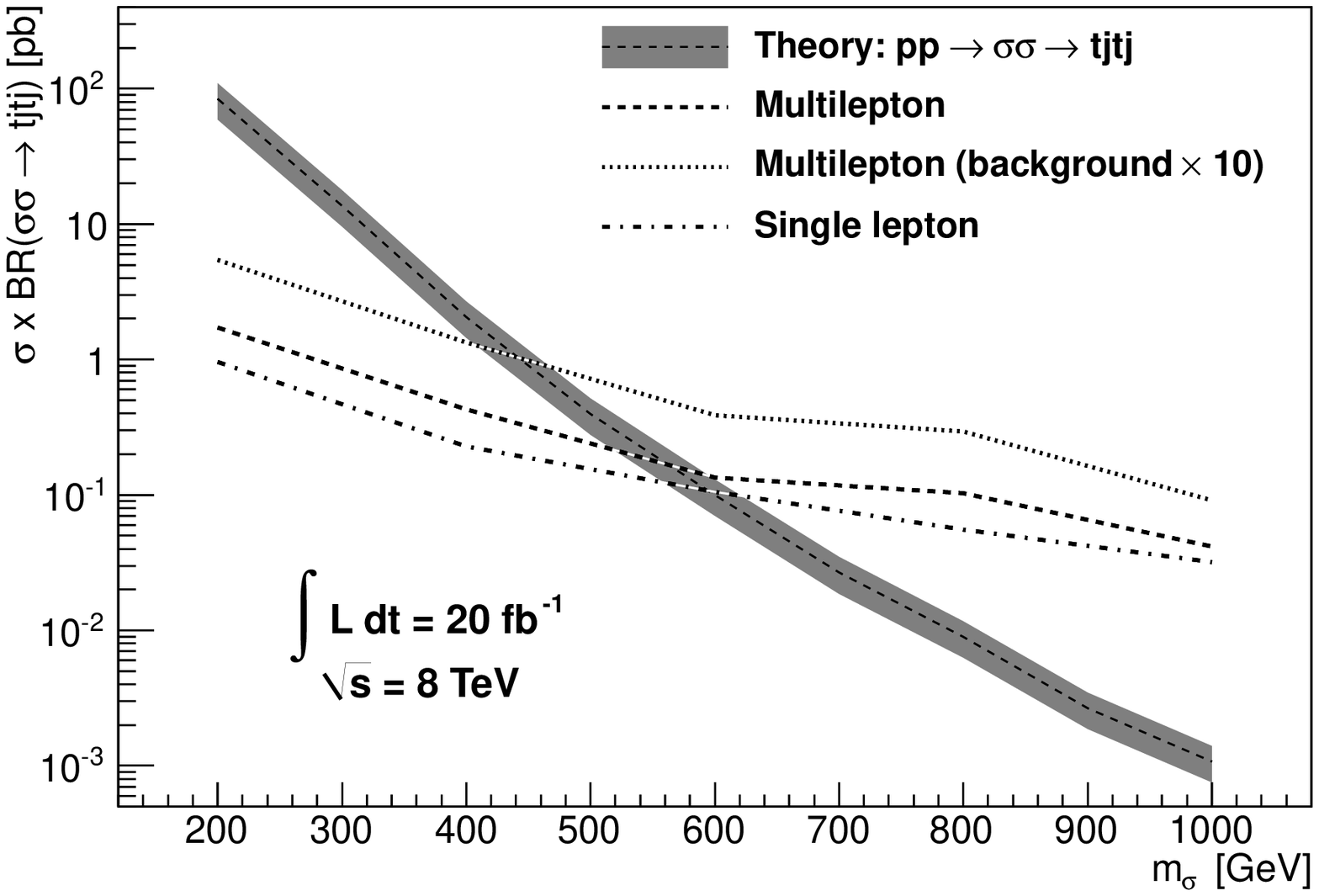} \\
\vspace{.1cm}
    \includegraphics[width=0.56\columnwidth]{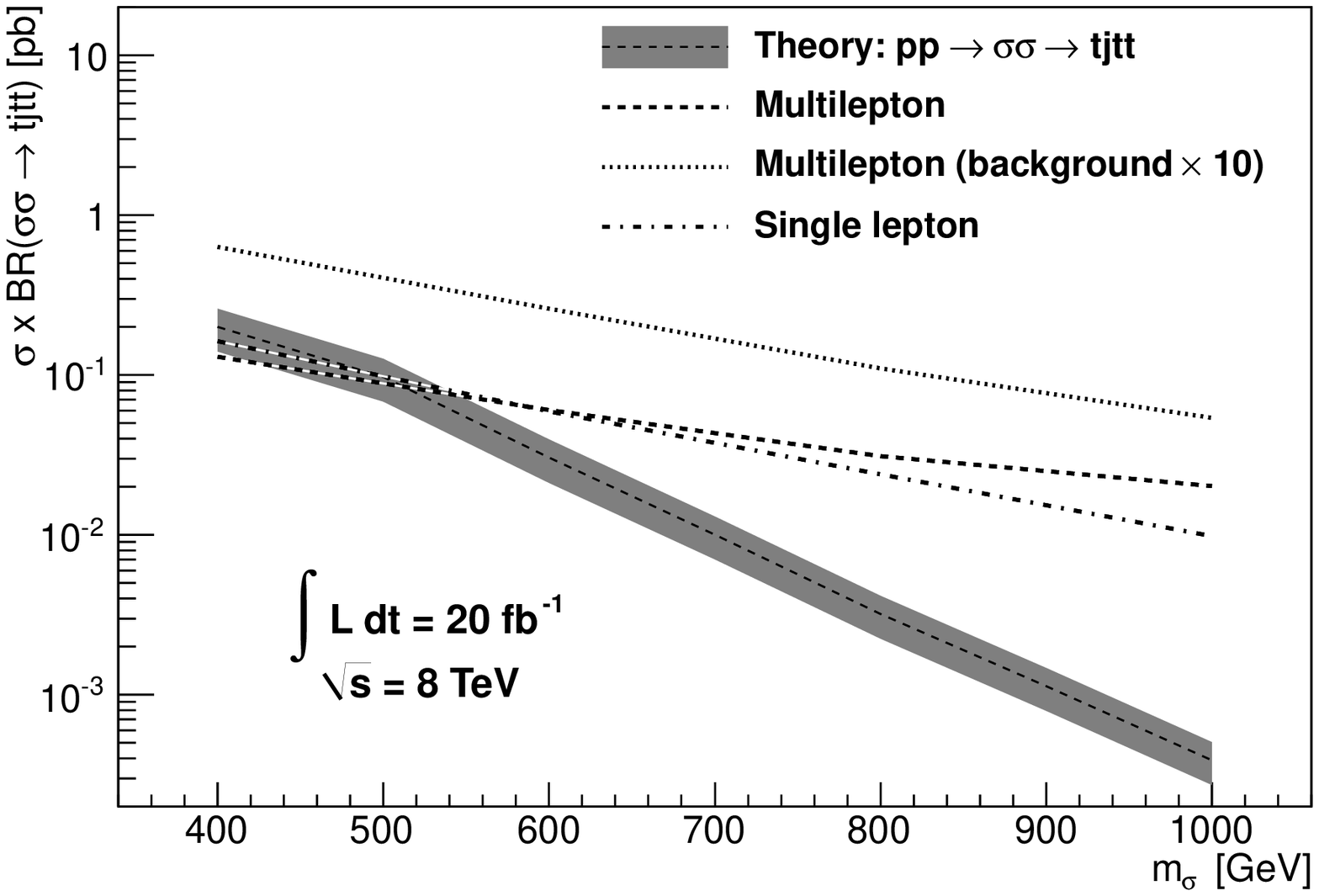} \\
\vspace{.1cm}
    \includegraphics[width=0.56\columnwidth]{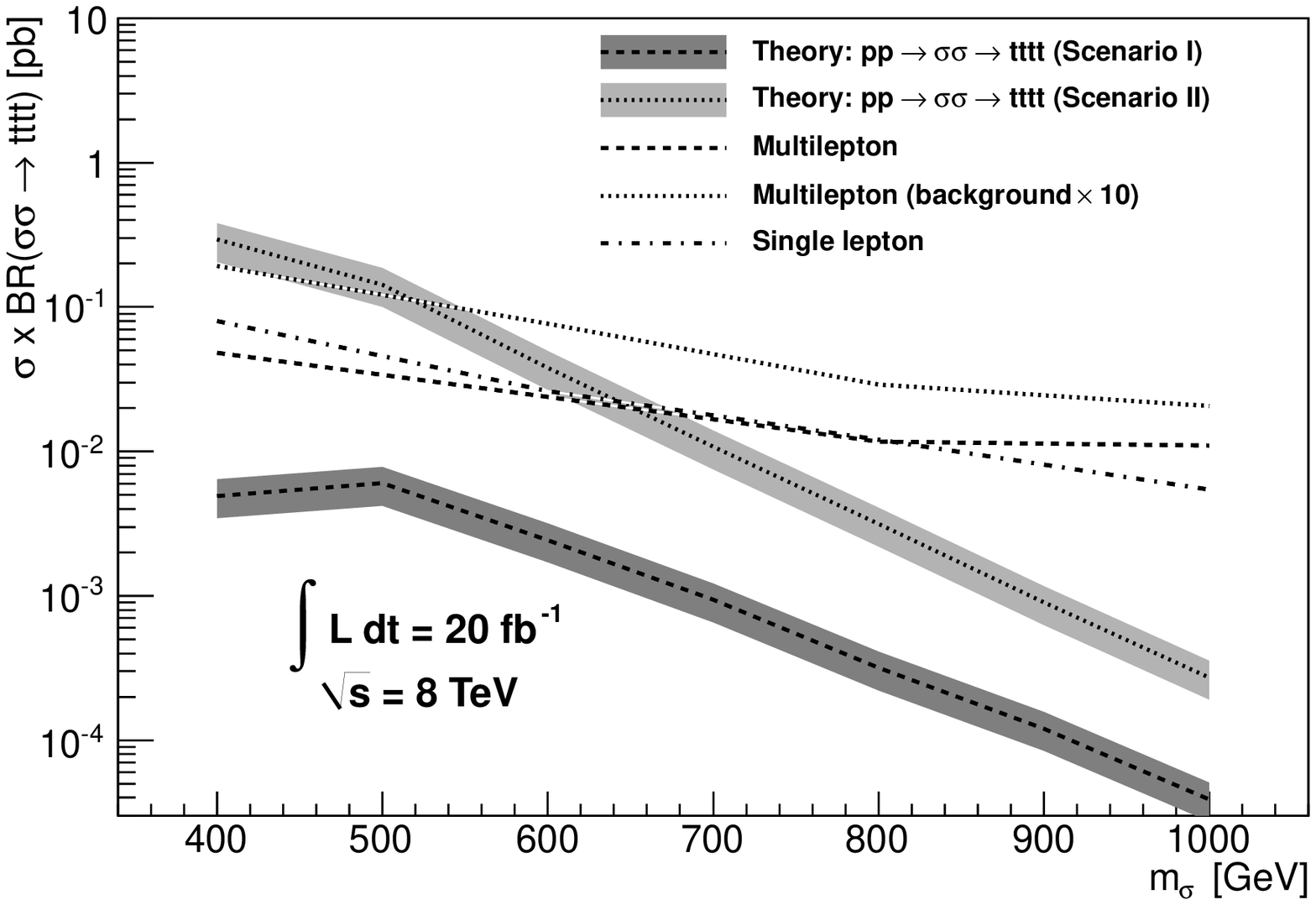} \\
    \caption{The 95\% confidence level expected signal cross sections as a
function of the
sgluon mass, for an integrated luminosity of 20 fb$^{-1}$ and $\sqrt{s}=8$ TeV.
The bands around the theoretical curves correspond to an uncertainty of 30\%,
and predictions for scenarios of class I (II) are presented
as light (dark) gray bands. The expected limits in the $tjtj$ (upper), $tjtt$
(middle) and $tttt$ channels (lower) are given, for the single lepton
(multilepton) analysis, as dot-dashed (dashed) curves. In the multilepton
case, we also show the results obtained after enhancing the background by a
factor of ten (dotted curves).}
    \label{fig:limits}
  \end{center}
\end{figure}

The results are presented in
Figure \ref{fig:limits} for the $tjtj$ channel (upper panel), $tjtt$ channel
(middle panel) and $tttt$ channel (lower panel) as dashed and dot-dashed curves
in the context of the multilepton and single lepton analysis, respectively. In
addition, the
effects of the non-simulated sources of background mentioned in Section
\ref{sec:multilepsel} are presented as dotted curves. On these figures, we also 
show the theoretical cross sections related to sgluon-induced
production of multitop final states as a function of the sgluon mass. 
In addition to the central NLO curves derived from 
Table \ref{tab:sigma}, we include a $30\%$ uncertainty band corresponding to the
typical order of magnitude of the variations of the results with respect to
different choices for the factorization and renormalization scales
\cite{GoncalvesNetto:2012nt}. The curves associated with scenarios of class I
are shown in dark gray, while those related to scenarios of class II are given
in light gray.

\renewcommand{\arraystretch}{1.2}
\begin{table}[t] 
  \begin{center}
    \begin{tabular}{ | c || c || c || c |}
     \hline 
      & \multirow{2}{*}{Single lepton analysis} & 
        \multirow{2}{*}{Multilepton analysis} & Multilepton analysis \\
      &                                         &
        & (background $\times 10$) \\
     \hline
      $tjtj$ & 590 $^{+40}_{-30}$ GeV & 570 $^{+30}_{-50}$ GeV  & 440
       $^{+40}_{-15}$ GeV \\
      $tjtt$ & 480 $^{+70}_{-80}$ GeV & 520 $^{+35}_{-90}$ GeV  & -  \\
      $tttt$ (Scenario I)  & - & -  & -\\
      $tttt$ (Scenario II) & 640 $^{+40}_{-30}$ GeV & 650 $^{+30}_{-40}$ GeV  &
         520 $^{+50}_{-110}$ GeV \\
     \hline 
    \end{tabular}
    \caption{Expected sensitivity of the LHC collider, running at a
center-of-mass of 8 TeV and for an integrated luminosity of 20 fb$^{-1}$, to
sgluons. The results are given, together with the associated
$1\sigma$ statistical uncertainties,
in terms of upper bounds on the sgluon mass to be reached, at the 95\%
confidence level, for different types of analyses and scenarios. 
For scenarios of class I, the expectations for the
$tjtj$ channel are given on the first line of the table, those related to the
$tjtt$ channel on its second line and those associated with the
$tttt$ channel on its third line. Scenarios of class II are only investigated in
the context of the $tttt$ search-channel and results are shown
on the fourth line of the table. Finally, we distinguish results obtained by
employing a single lepton
analysis (second column), a multilepton analysis after neglecting the QCD
background (third column) and a multilepton analysis after multiplying the
simulated background by a factor of ten (fourth column).} 
    \label{tab:recap_limits} 
  \end{center}
\end{table} 
\renewcommand{\arraystretch}{1.}

Sgluon masses excluded at the $95\%$ confidence level are also indicated in Table
\ref{tab:recap_limits}, for each final state and for each scenario considered in
this paper. The quoted uncertainties correspond to an estimation of the cross
section limits at the $1\sigma$ level, where $\sigma$ stands for the statistical
uncertainty. 

In the multilepton analysis, sgluon masses lower than 570 and 520 GeV can be
excluded in the $tjtj$ and $tjtt$ topologies, respectively, for scenarios of
class I. Equivalently, the ATLAS experiment is
sensitive to sgluon-induced multitop production cross sections of ${\cal
O}(100)$ fb for both the $tjtj$ and $tjtt$ signatures at $\sqrt{s}=8$ TeV.
Inspecting the second column of Table \ref{tab:recap_limits}, the limits
obtained for the $tjtj$ search-channel barely vary (by less than $10\%$) when
accounting for $1\sigma$ statistical uncertainties.
In contrast, the bounds extracted from the analysis of the $tjtt$ topology are
found to be more sensitive to statistics as they can vary by about 17\% with
respect to (un)lucky fluctuations. This feature can be understood from the
behavior of the 
theoretical cross sections in the 500-600 GeV sgluon mass range. While the
$tjtj$ cross section decreases with the sgluon mass, the $tjtt$ cross section is
rather flat. 
The multilepton analysis is however not sensitive to sgluon-induced production
of four top quarks for class I scenarios, at $\sqrt{s} =
8$ TeV and for an integrated luminosity of 20 fb$^{-1}$. This results from 
the too low branching ratio of the sgluon decay into a top-antitop pair (see Table
\ref{tab:sigma}). In class II scenarios, this branching ratio is $2.5-7.6$ times
more important, so that sgluon masses lower than 650 GeV, or cross section of
${\cal O}(10)$ fb, can be excluded at the 95\% confidence level. 

When taking into account non-simulated backgrounds, the total
number of expected background events is conservatively multiplied by ten.
Consequently, the $tjtt$ analysis is not sensitive to sgluon-pair production
anymore, while the masses excluded when analyzing $tjtj$ and $tttt$ (for
scenarios of class II) final states decrease from 570 to 440 GeV and 650 to 520
GeV, respectively.

We now turn to single lepton analyses and show that the reconstructed sgluon
mass from a $tjtj$ topology can be used to exclude, at the 95\% confidence
level, sgluon as heavy as 590 GeV (dot-dashed curve). Moreover, these bounds are
found not to drastically vary when including $1\sigma$ statistical uncertainties. 
Concerning the $tjtt$ and $tttt$ topologies, the $H_T$ variable is considered as a
discriminant and sgluon masses up to 480 GeV and 640 GeV can be reached. 
In the first case, statistical fluctuations can lead to different expectations
by about $\pm 15\%$ in the $tjtt$ case while in the $tttt$ case, the results are
found only to slightly change by about $5\%$.

\section{Conclusion}\label{sec:conclusions}
Many new physics theories predict the existence of a scalar
field, commonly dubbed sgluon, lying in the adjoint representation of the QCD
gauge group. To investigate the sensitivity of the LHC to this particle in the
case it couples dominantly to the top quark, as motivated by
hybrid $N\!=\!1 \!/\! N\!=\!2$ or $R$-symmetric supersymmetric theories, 
an effective field theory has been built. This theory consists of a minimal
extension of the Standard Model including a single real sgluon
field, together with a set of interactions leading to
its production and decay at the LHC. The model has been implemented in the
{\sc FeynRules} package so that a UFO model for {\sc MadGraph} has been
extracted. 

Final states containing two, three, or four top quarks produced in
association with jets have been investigated. To optimize the sensitivity
of the search for sgluon-pair production predominantly decaying to top quarks,
two signatures (multilepton and single lepton) have been considered in three
topologies ($tjtj$, $tjtt$, $tttt$). We have shown
that sgluons of about 500-700 GeV can be
reached at the LHC collider running at a center-of-mass energy of 8~TeV and for
an integrated luminosity of 20 fb$^{-1}$. 
Equivalently, this mass constraint can be translated into a bound
on the cross section associated with the production of two, three, or four top
quarks issued from the decays of a pair of sgluons. Contributions higher than
10-100 fb can be excluded for most scenarios, at the 95\% confidence level.
While a search strategy based on a single lepton selection ensures, with
appropriate selection criteria, the multijet background to be under good
control, non-simulated
contributions to the background such as jets faking leptons and
charge misidentification can reduce the sensitivity of the multilepton
signature by several hundreds of GeV.

Our work therefore motivates a future extension of the two performed analyses
in the context of a full detector simulation of the LHC experiments.

\acknowledgments
The authors are grateful to D.\ Zerwas for enlightening discussions at the
beginning of this project. Moreover, we thank the entire {\sc MadGolem} team,
and more in particular D.\
Gon\c calves Netto, D.\ L\'opez-Val and K.\ Mawatari, for helpful and interesting
discussions as well as for providing us next-to-leading order $K$-factors. This
work has been supported by the Theory-LHC-France initiative of the
CNRS/IN2P3 and a Ph.D.\ fellowship of the French ministry for education and
research. BF acknowledges partial support from the French ANR 12 JS05 002 01
BATS@LHC. 

\end{document}